\documentclass[twocolumn,showpacs]{revtex4}
\usepackage{graphicx}
\usepackage{dcolumn}
\usepackage{bm}
\usepackage{epsfig}

\newcommand{\be}{\begin{equation}}
\newcommand{\ee}{\end{equation}}
\newcommand{\ba}{\begin{eqnarray}}
\newcommand{\ea}{\end{eqnarray}}

\input{tcilatex}

\begin{document}

\title{Refractive effects in the scattering of loosely bound nuclei}
\author{Florin~Carstoiu$^{1,2,3}$, Livius~Trache$^{1}$, Robert~ E.~ Tribble$%
^{1}$, Carl A. Gagliardi$^{1}$ }
\affiliation{$^{1}$ Cyclotron Institute, Texas A\& M University, College Station, TX
77843-3366, USA \\
$^{2}$ Laboratoire de Physique Corpusculaire, IN2P3-CNRS, ISMRA et \\
Universit\'e de Caen, F-14050 Caen cedex, France \\
$^{3}$ National Institute for Physics and Nuclear Engineering "Horia
Hulubei', P.O. Box MG-6, 76900 Bucharest-Magurele, Romania }
\date{\today}

\begin{abstract}
A study of the interaction of loosely bound nuclei $^{6,7}$Li at 9 and 19 A
MeV with light targets has been undertaken. With the determination of
unambiguous optical potentials in mind, elastic data for four
projectile-target combinations and one neutron transfer reaction $^{13}$C($%
^{7}$Li,$^{8}$Li)$^{12}$C have been measured on a large angular range. The
kinematical regime encompasses a region where the mean field (optical
potential) has a marked variation with mass and energy, but turns out to be
sufficiently surface transparent to allow strong refractive effects to be
manifested in elastic scattering data at intermediate angles. The identified
exotic feature, a "plateau" in the angular distributions at intermediate
angles, is fully confirmed in four reaction channels and interpreted as a
pre-rainbow oscillation resulting from the interference of the barrier and
internal barrier far-side scattering subamplitudes.
\end{abstract}

\maketitle

%\date{December 26, 2003}

\section{\label{sec:intro}Introduction}

The study of nucleus-nucleus elastic scattering has a long history and
remains of interest due to both successes and failures that mark it (see for
example, Refs. \cite{satchler-love,brandan-s} and references therein). It is
an important subject \textit{per se}, and is also important as a tool for
the description of a series of phenomena that involve the distorted waves
given by optical model potentials (OMP). We are searching here for reliable
ways to predict optical model potentials for reactions with radioactive
nuclear beams (RNB). In particular our interest focuses on finding reliable
descriptions for transfer reactions involving relatively light, loosely
bound nuclei, which are used in indirect methods in nuclear astrophysics. A
range of RNB studies were made at energies around 10 MeV/nucleon, where the
reactions are peripheral, with the intent to obtain information about the
surface of the nuclei involved. These reactions use DWBA techniques to
extract nuclear structure information. However, the well known existence of
many ambiguities in the OMPs extracted from elastic scattering can raise
questions about the accuracy of these determinations. Therefore, we are
searching for ways to reduce these ambiguities and to predict OMP for
reactions with RNBs. Experimental studies using RNBs have, heretofore, not
been suitable for detailed elastic scattering analyses. The closest we can
get using stable beams is by studying the elastic scattering of loosely
bound nuclei. We chose here to study the elastic scattering of $^{6,7}$Li
projectiles, because they are fragile (loosely bound), with a pronounced
cluster structure and with low Z and can, therefore, exhibit a range of
phenomena, involving absorption, diffraction and refraction, mostly of
nuclear nature.

Earlier we have carried out a study of elastic scattering around 10A MeV for
a range of projectile-target combinations involving $p$-shell nuclei \cite%
{trache00}. We found a relatively simple method to predict OMP for loosely
bound nuclei, based on the renormalization of the independent real and
imaginary terms obtained from a double folding procedure using the JLM
nucleon-nucleon (NN) effective interaction. The procedure successfully
described the data for all the projectile-target combinations and the
energies in the study for most of the angular ranges measured. In one single
case ($^{7}$Li at 99 and 130 MeV on $^{12}$C) the folding potentials failed
to describe well the large angle data. Later, the results were used to
describe elastic scattering angular distributions measured in a series of
experiments with RNBs at or around 10 MeV/nucleon: $^{7}$Be on $^{10}$B and
melamine targets \cite{azhari99}, $^{11}$C \cite{tang03}, $^{13}$N \cite%
{tang04} and $^{17}$F \cite{blackmon03} on $^{12}$C and $^{14}$N targets. We
return to that study here with new data extending the angular ranges for the 
$^{7}$Li scattering and adding data for $^{6}$Li scattering and with a
refined analysis.

Recent work \cite{oglob,szilner} has established that elastic scattering of
light tightly bound heavy ion systems such as $^{16}$O+$^{12}$C and $^{16}$O+%
$^{16}$O show sufficient transparency for the cross section to be dominated
by the far-side scattering. Intermediate angle structures appearing in the
elastic scattering distributions at angles beyond the Fraunhofer diffractive
region have been identified as Airy minima of a nuclear rainbow, i.e. a
destructive interference between two far-side trajectories which sample the
interior of the potential. A number of high order Airy minima have been
identified by observing that such structures are largely insensitive to an
artificial reduction of the absorption in the optical potential, and
therefore they appear as a manifestation of the refractive power of the
nuclear potential. While at high energy \cite{stil} this picture was well
substantiated by a semiclassical nonuniform decomposition of the scattering
function \cite{knoll}, at lower energies the situation is more difficult to
understand. It has been shown by Anni \cite{anni}, that such structures
could be explained by the interference of two amplitudes appearing in
different terms of a multireflection uniform series expansion of the
scattering amplitude and therefore the interpretation using rainbow
terminology is not appropriate.

For loosely bound nuclei the situation is even more uncertain. When a
nucleon or a group of nucleons has small separation energy, the wave
function penetrates well beyond the potential range. The corresponding
components in the optical potential are expected to be more diffuse as
compared to normal nuclei, leading to a competition between the increased
refractive power of the real potential and the increased absorption at the
nuclear surface. The small separation energy implies also that the dynamic
polarization potential (DPP) \cite{feshbach} arising from the coupling to
breakup states may be strong and with a complicated energy and radial
dependence. It follows that for loosely bound nuclei the DPP cannot be
treated as a small perturbation and the usual phenomenological procedure in
renormalizing the folding potential form factor may be questioned. It has
been estimated that the DPP is strongly repulsive at the nuclear surface in
the case of $^{6}$Li \cite{sakur} and this prompted Mahaux, Ngo and Satchler 
\cite{mahaux} to conjecture that for loosely bound nuclei the barrier
anomaly may be absent due to the cancellation between the repulsive (DPP)
and attractive (dispersive) components of the optical potential.

In the specific case of $^{6,7}$Li scattering on light targets a large body
of data have been accumulated in the range 5-50 MeV/nucleon. At high energy,
Nadasen and his group \cite{nadas1,nadas2} have been able to derive a unique
optical potential which was essential to assess the quality of the folding
model. At lower energies, ambiguities found in the analysis of data
prevented any definite conclusion about the strength and energy dependence
of the optical potential. A study by Trcka \textit{et al.}\cite{trcka} on $%
^{6}$Li+$^{12}$C elastic scattering at 50 MeV, found an exotic feature
("plateau") in the angular distribution of the elastic scattering at
intermediate angles which resembles similar structures found in more bound
systems. They interpreted the structure as a diffractive effect arising from
an angular momentum dependent absorption. There are experimental hints that
such structures also\ appear in neighboring systems, $^{6}$Li+$^{16}$O and $%
^{6}$Li+$^{9}$Be, as a possible manifestation of the average properties of
the interaction potential.

In this paper we present a precision measurement of elastic scattering of $%
^{6,7}$Li on $^{12,13}$C and $^{9}$Be targets at 9 and 19 MeV/nucleon. The
lower energy was chosen in view of our systematic studies of nuclear
reactions for astrophysics. The higher energy is close to the saturation
energy for these projectiles, i.e. the energy where almost all reaction
channels are open. The "plateau" feature is confirmed in four
projectile-target combinations at 9 MeV/nucleon. The high selectivity
induced by this structure allowed the derivation of an almost unique
Woods-Saxon optical potential. A folding model analysis using the complex,
density and energy dependent NN interaction of Jeukenne, Lejeune and Mahaux
(JLM) \cite{jeuken}, where corrections due to the strong DPP have been
included, confirmed that our elastic distributions could be described using
deep and extremely transparent potentials. The remaining ambiguities have
been eliminated using an accurate dispersion relation analysis. The
intermediate angle structures have been discussed using the semiclassical
uniform approximation for the scattering function of Brink and Takigawa \cite%
{brink2}. We explain the intermediate angle structure as a coherent
interference effect of two subamplitudes corresponding to trajectories
reflected at the barrier and interfering with trajectories which sample the
nuclear interior. Thus, this refractive effect appears as a signature of a
highly transparent interaction potential.

The paper is structured in the following way: after this introduction, the
experimental methods are discussed in Sect. \ref{sec2}, the analysis of the
elastic scattering data using phenomenological and microscopic optical model
potentials is discussed in Sect. \ref{sec3}, and the implications of this
analysis for the transfer reaction ($^{7}$Li,$^{8}$Li) is discussed in Sect. %
\ref{sec4}. In Sect. \ref{sec5} the dispersion relation is used to put
additional constrains on the potentials extracted, followed by a discussion
of the decomposition of the far-side scattering amplitude into barrier and
internal barrier components responsible for the "plateau" structure at
intermediate angles (Sect. \ref{sec6}), and the conclusions (Sect. \ref%
{concl}).

%%%%%%%%%%%%%%%%%%%%%%%%%%%%%%%%%%%%%%%%%%%%%%%%%%%%%%%%%%%%%%%%%%%%%

\section{The experiments}

\label{sec2} The experiments were performed using $^{6}$Li and $^{7}$Li
beams of 9 and 19 MeV/nucleon from the Texas A\&M University K500
superconducting cyclotron and the Multipole Dipole Multipole (MDM) magnetic
spectrometer \cite{Oxs}. A list of the measurements is given in Table \ref%
{tabexp}. The measurements with $^{7}$Li were done to extend the angular
range that was effectively covered in earlier work. The experimental setup
and the data reduction procedures were similar to those used in Ref. \cite%
{trache00}. The beams were prepared using the beam analysis system \cite{BAS}%
, which allows for the control of the energy spread ($\Delta E/E$ up to
1/2500) and angular spread ($0.1^{\circ }$) of the beam. Self-supported $%
^{9} ${Be }(200 $\mu $g/cm$^{2}$ thick){, }$^{{12}}${C }(260 $\mu $g/cm$^{2}$%
) and $^{13}$C (390 $\mu $g/cm$^{2}$) targets were placed perpendicular to
the beam in the target chamber of the MDM. The magnetic field of the MDM
spectrometer was set to transport the fully stripped Li ions to the focal
plane where they were observed in the modified Oxford detector \cite{Oxd}.
In the detector, the position of the particles along the dispersive
direction was measured with resistive wires at four different depths,
separated by about 16 cm each. For particle identification we used the
specific energy loss measured in the ionization chamber and the residual
energy measured in a NE102A plastic scintillator located behind the output
window of the detector. The input and output windows of the detector were
made of 1.8 and 7.2 mg/cm$^{2}$ thick Kapton foils, respectively. The
ionization chamber was filled with pure isobutane at $40$ torr. The entire
horizontal acceptance of the spectrometer, $\Delta \theta =\pm 2^{\circ }$,
and a restricted vertical opening, $\Delta \phi =\pm 0.5^{\circ }$, were
used in the measurements at forward angles, whereas at the largest angles
the vertical opening of the acceptance window was raised to\ $\Delta \phi
=\pm 1.0^{\circ }$. Raytracing was used to reconstruct the scattering angle.
For this purpose, position calibration of the detector was performed by
using the scattering on a thin Au target (212 $\mu $g/cm$^{2}$) and an angle
mask consisting of five openings of $\delta \theta $=0.1$^{\circ },$ located
at -1.6$^{\circ }$, -0.8$^{\circ }$, 0$^{\circ }$, +0.8$^{\circ }$ and +1.6$%
^{\circ }$ relative to the central angle of the spectrometer. In addition to
RAYTRACE \cite{RAY} calculations, angle calibration data were obtained at
several angles by using the angle mask. Typically the spectrometer was moved
by $2^{\circ }$ or $3^{\circ }$ at a time, allowing for an angle overlap
that provided a self-consistency check of the data. Normalization of the
data was done using current integration in a Faraday cup. Focal plane
reconstruction was done at each angle using the position measured with the
signals in the wire nearest to the focal plane and using the detector angle
obtained from the position measured at two of the four wires (typically the
first and last). The angular range, $\Delta \theta =4^{\circ }$, covered by
the acceptance slit was divided into 8 bins, resulting in 8 points in the
angular distribution being measured simultaneously, with each integrating
over $\delta \theta _{lab}=0.5^{\circ }$.

The measurements with the angle mask showed that the resolution in the
scattering angle (laboratory) was $\Delta \theta _{res}=0.18^{\circ
}-0.25^{\circ }$ full-width at half maximum (FWHM). This includes a
contribution from the angular spread of the beam of about $0.1^{\circ }$.
The best energy resolution obtained at forward angles was 150 keV FWHM. It
degraded as we advanced to larger angles due to the large kinematic factor, $%
k=\frac{1}{p}\frac{dp}{d\theta }$, coupled with the finite angular spread in
the beam. However, it never degraded our ability to isolate the elastic
peak, even in the case of the $^{7}$Li experiments where the first excited
state of the projectile is only 477 keV away. The active length of the focal
plane allowed us to cover a total excitation energy of about 7 MeV, centered
around the elastic peak. Thus we were able to measure inelastic scattering
to the lowest excited states of the projectile-target systems at the same
time. These inelastic scattering data were used as additional information to
check the experimental procedures. In one of the runs we have also measured
the neutron transfer reaction $^{13}$C($^{7}$Li,$^{8}$Li)$^{12}$C at E($^{7}$%
Li)=63 MeV, which was discussed elsewhere in detail \cite{trache03} and is
used here to check the sensitivity of observables in other channels to the
OMP extracted from the elastic scattering data.

To obtain accurate absolute values for the cross sections, target thickness
and charge collection factors were determined by a two-target method as
described in Ref. \cite{mukh97}. We also determined the target thickness by
measuring the energy loss of alpha particles from a $^{228}$Th source and
the accuracy in normalization is 9\%. Combining the results of these
independent determinations, we conclude that we have an overall
normalization accuracy of 7\% for the absolute values of the cross sections.

%%%%%%%%%%%%%%%%%%%%%%%%%%%%%%%%%%%%%%%%%%%%%%%%%%%%%%%%%%%%%%%%%%%%%%%%%
%%%%%
%%

\section{Optical-model analysis}

\label{sec3}

\begin{figure*}[tbh]
\begin{center}
\mbox{\epsfig{file=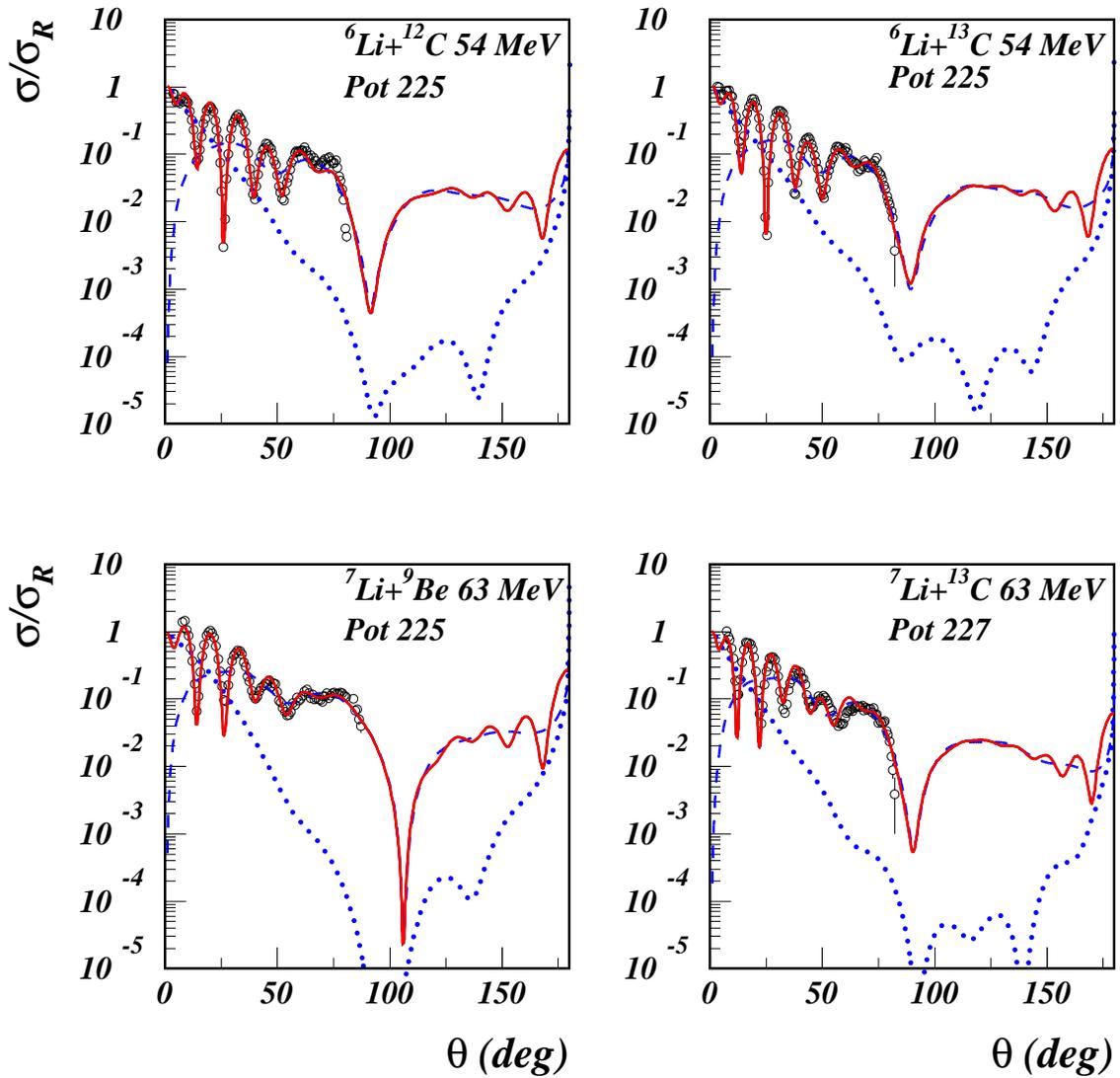,width=15cm}}
\end{center}
\caption{(Color online) Woods-Saxon optical model analysis (full lines) of
elastic scattering data (open points) at 9 MeV/nucleon (Table \protect\ref%
{tab1:res_ws}). Far-side/near-side cross sections are also shown by dashed
and dotted lines, respectively. The depth of the real potential is shown to
identify the particular WS potential parameters used in the calculations.}
\label{fig:ws}
\end{figure*}

The measured elastic data at 9 MeV/nucleon, shown in Fig. \ref{fig:ws} as
the ratio to the Rutherford cross section, extend to a larger angular range
than previously measured \cite{trache00}. These data show complex forms with
characteristic rapid oscillations at small angles followed by a marked
change in shape at intermediate angles: a plateau is developed at $\theta
=50^{\circ }-70^{\circ }$ followed by a deep minimum at $\theta \approx
80^{\circ }$. Assuming pure Fraunhofer scattering at forward angles, we
extract a grazing angular momentum $l_{g}\approx $15 from the angular
spacing $\Delta \theta =\pi /(l_{g}+1/2)$. %fig 1
The striking fact is that the same pattern emerges for all four
projectile-target combinations, including that for the $^{9}$Be target where
a much stronger absorption is expected. (We remind the reader that $^{9}$Be
is a perfect black disc target since it has very low thresholds for breakup
into the neutron and alpha channels $S_{n}$=1.66 MeV and $S_{\alpha }$=2.47
MeV, and there are no bound excited states. These values should be compared
with $S_{\alpha }$=1.47 MeV in $^{6}$Li and $S_{\alpha }$=2.47 MeV in $^{7}$%
Li.)

Similarities seen in the differential cross sections shown in Fig. \ref%
{fig:ws} indicate general wave-mechanical characteristics of the scattering
process and average systematic properties of the nuclear interaction.
Specific structure effects can be isolated only as small deviations from the
normal behavior. Therefore the data are analyzed using optical potentials
with conventional Woods-Saxon (WS) form factors for the nuclear term,
supplemented with a Coulomb potential generated by a uniform charge
distribution with a reduced radius fixed to $r_{c}$=1 fm. No preference has
been found for volume or surface localized absorption and throughout the
paper only volume absorption is considered. In the absence of any spin
dependent observables, spin-orbit or tensor interactions have been ignored.
Ground state reorientation couplings also have been neglected. The potential
is defined by six parameters specifying the depth and geometry of the real
and imaginary terms 
\begin{equation}
\mathrm{U(r)}=-\left( \mathrm{Vf_{V}(r)+iWf_{W}(r)}\right)  \label{p1}
\end{equation}%
where 
\begin{equation}
\mathrm{f_{x}(r)=\left[ 1+\exp \left( \frac{r-r_{x}(A_{1}^{1/3}+A_{2}^{1/3})%
}{a_{x}}\right) \right] ^{-1}}  \label{p2}
\end{equation}%
and x=V,W stands for the real and imaginary parts of the potentials,
respectively. The number of data points per angular distribution exceeds
N=100 points and therefore the usual goodness of fit criteria ($\chi ^{2}$)
normalized to N has been used. A source of bias was the finite angular
acceptance of the detectors (the 0.5$^{\circ }$ bins, in the present case).
The averaging associated with this finite angular resolutions has most
effect on the depth of sharp minima. A few exploratory calculations showed
that allowing the normalization to vary did not result in any qualitative
changes and did not indicate that any renormalization by more than a few
percent would be preferred. Optical parameter sets collected from the
literature were used as starting values for the search procedure. In
particular the potential OM1 of Trcka \textit{et al.} \cite{trcka} has been
extensively tested. Guided by these potentials and by our earlier analysis 
\cite{trache00} a number of some $10^{6}$ potentials with real volume
integrals in the range $J_{V}=200-600$ MeV fm$^{3}$ have been generated for
each reaction channel, thus exploring the functional Woods-Saxon space in
full detail. Local minima were identified and a complete search on all six
parameters determined the best fit potentials. The plateau feature at the
intermediate angles and the sharp decrease in the cross section near $\theta
=80^{\circ }$ could be fitted only with deep potentials with real volume
integrals (per nucleon) exceeding a critical value $J_{Vcrit}\approx $ 300
MeV fm$^{3}$. There is a consistent preference for potentials with
relatively weak imaginary parts with values of $W$ around 15 MeV except for $%
^{7}$Li scattering where somewhat larger values are needed to fit the data.
We systematically find $r_{V}<r_{W}$ and large diffuseness parameters $%
a_{V}\simeq a_{w}\simeq $ 0.8 fm in agreement with theoretical expectations
for loosely bound nuclei \cite{hussein,bonacc}. A grid search procedure on
the real depth of the potential allowed to identify discrete ambiguities.
Parameters for the first two discrete families are given in Table \ref%
{tab1:res_ws}. These are identified by a jump of $\Delta J_{V}\approx $ 100
MeV fm$^{3}$ from one family to another and almost constant imaginary volume
integral. As a consequence, the total reaction cross section seems to be a
well determined observable. Gridding on other WS parameters revealed a
continuous ambiguity of the form $J_{V}R_{V}\approx $ const, where $R_{V}$
is the $rms$ radius of the potential. The larger the volume integral, the
smaller the radius that is required to fit the data. This is a clear
manifestation of a complicated radial dependence of the dynamic polarization
potential (DPP) which may lead to radii much smaller than the minimal value
implied by the folding model (e.g., $R_{F}^{2}=R_{1}^{2}+R_{2}^{2}$, for a
zero range NN effective interaction). However, for each discrete family
rather precise values of the $rms$ radii were required to fit both forward
and intermediate angle cross sections.

Sometimes more subjective criteria may be used to choose between various
ambiguous potentials based upon general theoretical expectations. For
example one may require consistency with the results of analyses of other
data for the same system at nearby energies with the expectation that the
potential should not change rapidly with mass and energy. Individual elastic
data sets possess individual idiosyncrasies which facilitate the inference
of a single local potential. We note that, seemingly, there is a
compatibility between all data sets: an optimum potential found for one data
set gives already a good fit to the other. In fact, potentials given as
first entry in Table \ref{tab1:res_ws} were obtained by iterating several
times this procedure in an attempt to find a single potential which would
simultaneously fit all data at 9 MeV/nucleon. A compromise could be obtained
with transparent deep potentials close to $V_{0}\approx $ 225 MeV having a
strongly refractive core at small radii, surrounded by a weakly absorptive
halo. In fact, examining the ratio $w(r)=W(r)/V(r)$ \cite{brandan} as a
function of the radial distance, we found that our potential may by
qualified as having internal ($r\sim 0-4$ fm) and surface ($r>8$ fm)
transparency ($w\approx 0.1$) but with a pronounced maximum ($w\approx 0.8$)
near the empirical strong absorption radius ($R_{s}\approx 6$ fm) in
agreement with the systematics found in other more bound systems \cite%
{brandan}. The surface localized absorption suggests that the reaction
mechanism is dominated by direct reactions. The relatively large radius of
the absorption required by the data is an indication that fusion already
sets in the region of the barrier and that fusion is a large component of
the total reaction cross section.

\begin{figure}[tbh]
\begin{center}
\mbox{\epsfig{file=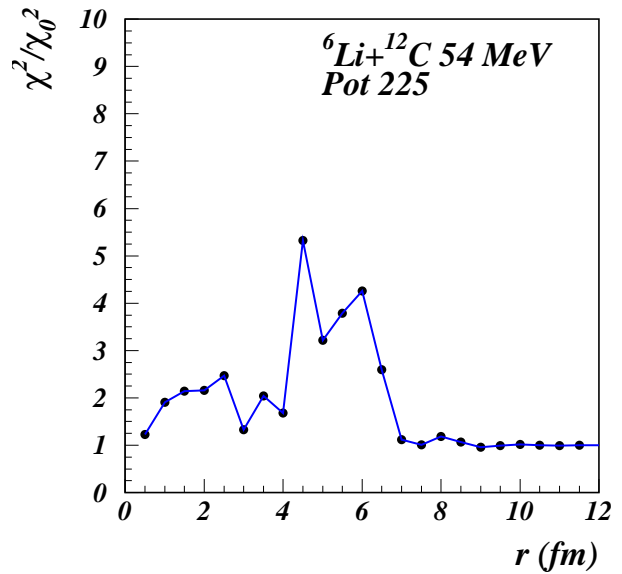,width=8.cm}}
\end{center}
\caption{(Color online) Notch test for the radial sensitivity of the
potential. The bare $\protect\chi _{0}^{2}$ is calculated for the first
potential in Table \protect\ref{tab1:res_ws}. A Gaussian spike with a 10$\%$
strength and a width (FWHM) of 0.2 fm is superimposed on the bare potential
at various radii. Points are connected by lines to guide the eye.}
\label{fig:spike}
\end{figure}

A variety of notch tests have been performed to determine the radial
sensitivity of the potential. One test was done using a Gaussian spike
superimposed on the real potential at a given radius. The resulting
influence on the $\chi ^{2}$ of the fits is displayed in Fig. \ref{fig:spike}%
. It shows that there is a relatively high sensitivity for radial distances
as low as 4-6 fm, well inside the strong absorption radius. Deeper inside
this radial range, the refractive index, defined as $n=\sqrt{1-\frac{V}{%
E_{cm}}}$ is almost real and reaches values as high as $n=2.6$, comparable
to that of diamond. %fig 2

\begin{figure}[tbh]
\begin{center}
\mbox{\epsfig{file=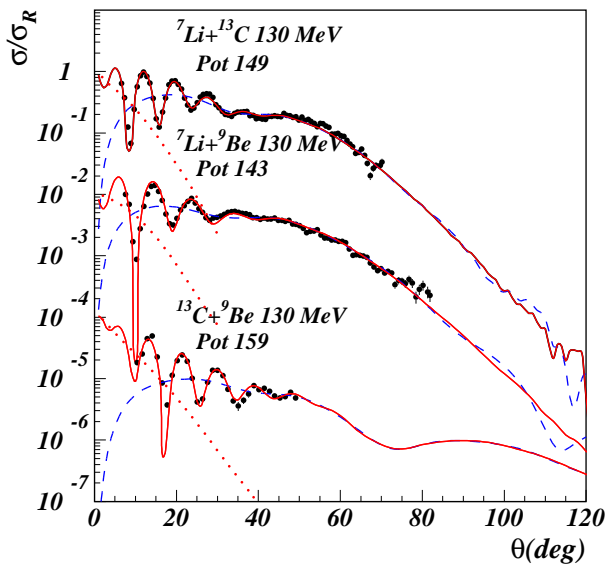,width=8cm}}
\end{center}
\caption{(Color online) Woods-Saxon optical model analysis of elastic data
at 130 MeV ( Table \protect\ref{tab1:res_ws} ). Far-side/near-side cross
sections are also shown by dashed and dotted lines, respectively. The lower
curves are multiplied by $10^{-2}$ and $10^{-4}$, respectively.}
\label{fig:ws130}
\end{figure}

As mentioned already, it was shown in Refs. \cite{oglob,szilner,stil} that
the elastic scattering of light heavy ion systems such as $^{16}$O+$^{12}$C
and $^{16}$O+ $^{16}$O shows sufficient transparency for the cross section
to be dominated by far-side scattering. Structures appearing in the elastic
scattering angular distributions at intermediate angles have been identified
as Airy minima of a nuclear rainbow, due to a destructive interference
between two far-side trajectories which sample the interior of the
potential. At 19 MeV/nucleon the $^{7}$Li scattering data show rapid,
diffractive Fraunhofer oscillations at small angles due to the strong
near-far amplitude interference (Fig. \ref{fig:ws130}). Beyond the crossover
the near-side amplitude makes a negligible contribution to the cross
section. The shoulder and the deep minimum seen at 9 MeV/nucleon (Fig. \ref%
{fig:ws}) are washed out in the far-side amplitude and only a broad, less
pronounced minimum survives, followed by a broad Airy maximum and an
exponential, structureless decay of the cross section at large angles.
Clearly, both the data at 9 and 19 MeV/nucleon (Figs. \ref{fig:ws} and \ref%
{fig:ws130}) show far-side dominance as a possible manifestation of
refractive effects. However, this simple dominance does not explain, by
itself, the difference in the angular distributions seen at these energies,
suggesting a difference in the reaction mechanism. In fact the above picture
has been already challenged by Anni \cite{anni} and by Michel \textit{et al.}
\cite{miche} for the simple reason that the far-side amplitude has never
been decomposed in subamplitudes which would explain the quoted
interference. We come back to this topic in Section \ref{sec6}. For the
moment we adopt the interpretation of Michel \textit{et al.} \cite{miche}
and denote the complex structure at intermediate angles in our data as
pre-rainbow oscillations. %fig 3

In the remainder of this section we discuss the ability of the folding model
to describe the pre-rainbow oscillation seen at 9 MeV/nucleon and the
rainbow patterns at higher energies. Data at somewhat lower energies are
also examined in order to see if the plateau feature persists in adjacent
systems and on a larger energy range. Our preferred model is the nuclear
matter approach of Jeukenne, Lejeune and Mahaux (JLM) \cite{jeuken} which
incorporates a complex, energy and density dependent parametrization of the
NN effective interaction obtained in a Brueckner Hartree-Fock approximation
from the Reid soft core nucleon-nucleon potential. The systematic study \cite%
{trache00} of the elastic scattering between $p$-shell nuclei at energies
around 10 MeV/nucleon led to the surprising result that on average, the
imaginary part of the folded JLM potential was perfectly adequate to
describe such reactions and did not need any renormalization ($N_{W}=1.00\pm
0.09$), while the real component needed a strong renormalization, in line
with other effective interactions used in folding models. However, the
present data extend to a much larger angular range and need further
refinements of this model.

In the JLM model the complex form factor for the optical potential is given
by 
\begin{equation}
U(R)=\int d\vec{r}_{1}d\vec{r}_{2}\rho _{1}(r_{1})\rho _{2}(r_{2})v(\rho
,E,s)g(s)  \label{eqjlm1}
\end{equation}%
where $v$ is the (complex) NN interaction, $\rho _{1(2)}$ are the single
particle densities of the interacting partners, calculated in a standard
spherical Hartree-Fock procedure using the energy density functional of
Beiner and Lombard with the surface term adjusted to reproduce the total
binding energy \cite{beiner,carstoiu-lomb}, $\vec{s}=\vec{r}_{1}+\vec{R}-%
\vec{r}_{2}$ is the NN separation distance between interacting nucleons and $%
\rho $ is the overlap density. The effective NN interaction contains an
isovector component which gives a negligibly small contribution for $p$%
-shell nuclei but is included here for convenience in conjunction with
appropriate single particle isovector densities. The smearing function $g(s)$
is taken as a normalized Gaussian \cite{jeuken,bauge,trache00}, 
\begin{equation}
g(s)=\frac{1}{t^{3}\pi ^{3/2}}\exp (-s^{2}/t^{2})  \label{eqjlm2}
\end{equation}%
which tends to a $\delta $-function for $t\rightarrow 0$, while for finite
values of the range parameter $t$ it increases the $rms$ radius of the
folding form factor by $r_{g}^{2}=(3/2)t^{2}$, leaving unchanged the volume
integral. Inclusion of a smearing function with a varying range parameter,
greatly increases the ability of the folding form factor to simulate the
radial dependence of DPP.

The geometric or arithmetic mean of the overlapping densities has been used
to define the overlap density $\rho $ in Eq. \ref{eqjlm1} 
\begin{equation}
\rho =[\rho _{1}(\vec{r}_{1}+\frac{1}{2}\vec{s})\rho _{2}(\vec{r}_{2}-\frac{1%
}{2}\vec{s})]^{1/2}  \label{eqjlm3}
\end{equation}%
and 
\begin{equation}
\rho =\frac{1}{2}[\rho _{1}(\vec{r}_{1}+\frac{1}{2}\vec{s})+\rho _{2}(\vec{r}%
_{2}-\frac{1}{2}\vec{s})].  \label{eqjlm4}
\end{equation}%
The former was introduced by Campi and Sprung in density-dependent
Hartree-Fock calculations \cite{campi}. It is physically appealing since the
overlap density tends to zero when one of the interacting nucleons is far
from the bulk, and to the nuclear matter saturation value at complete
overlap. The approximation in Eq. \ref{eqjlm4} is similar to that used in
folding calculations with density-dependent M3Y effective interactions \cite%
{khoa}, except for the factor $1/2$ which has been introduced here because
JLM interaction is defined only up to the nuclear matter saturation value $%
\rho \leq \rho _{0}$. It has been suggested to us \cite{oertzen} that the
drawbacks seen in our earlier analysis of the scattering of $^{7}$Li at 19
MeV/nucleon (see Fig. 6a in Ref. \cite{trache00}) may be due to the weak
density dependence introduced by Eq. \ref{eqjlm4}, and thus rainbow patterns
could not be reproduced. However, the optical model analysis presented above
showed clearly that the pre-rainbow oscillations (at 9 MeV/nucleon) and
rainbow patterns (at 19 MeV/nucleon) could be described if and only if the
potentials have the proper $rms$ radius. It turns out that the smearing
procedure described above is essential in simulating the complicated radial
dependence of the dynamic polarization potential. %fig4
\begin{figure*}[tb]
\begin{center}
\mbox{\epsfig{file=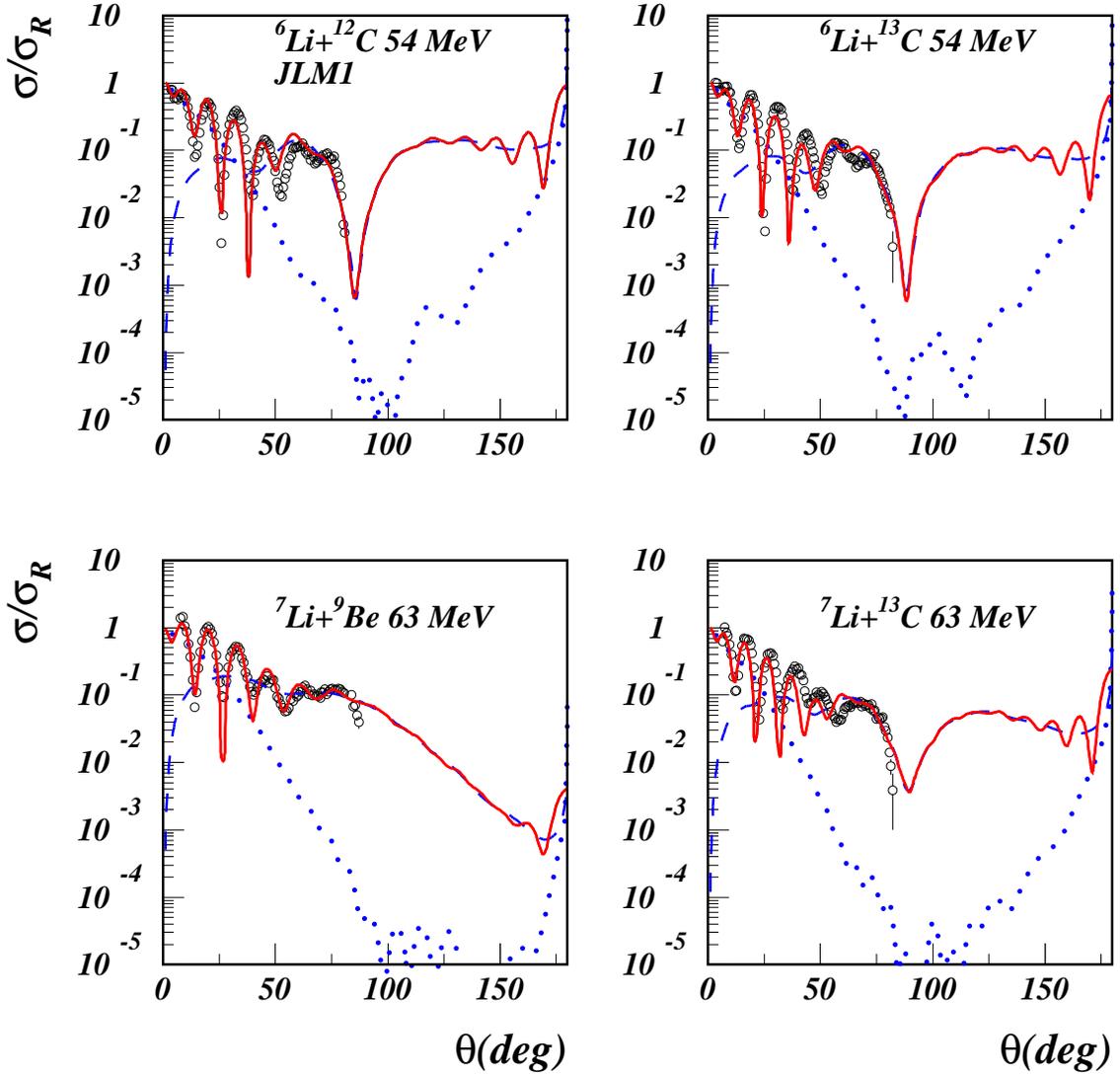,width=15cm}}
\end{center}
\caption{(Color online) Comparison of the JLM1 folding model calculations
(full lines) with present data at 9 MeV/nucleon. The parameters are given in
Table \protect\ref{tab1:res_JLM1}. Far-side (dashed) and near-side (dotted)
cross sections are indicated in the form of ratios to the Rutherford cross
sections. }
\label{fig:jlm1_1}
\end{figure*}

In the earlier analysis \cite{trache00}, fixed values for the range
parameters $t_{V}=1.2$ fm and $t_{W}=1.75$ fm, found from a global analysis
of the data were used. Only the renormalization factors $N_{V}$ and $N_{W}$
were left free in the fits for each case. In the present analysis with
double folded potentials, all four parameters: two strength parameters ($%
N_{V}$ and $N_{W}$) and two range parameters ($t_{V}$ and $t_{W}$), have
been searched simultaneously to fit the data for each case 
\begin{equation}
\mathrm{U}_{\mathrm{DF}}\mathrm{(r)}=\mathrm{N}_{\mathrm{V}}\mathrm{V(r,t}_{%
\mathrm{V}}\mathrm{)+iN}_{\mathrm{W}}\mathrm{W(r,t}_{\mathrm{W}}\mathrm{)}
\label{fp1}
\end{equation}%
to obtain a phenomenological representation of the DPP as a uniform
renormalization of the depths and radii of the folding potentials. The
calculations using Eqs. (\ref{eqjlm3}) and (\ref{eqjlm4}) are dubbed JLM1
and JLM2 respectively. %fig 5
\begin{figure*}[tbh]
\begin{center}
\mbox{\epsfig{file=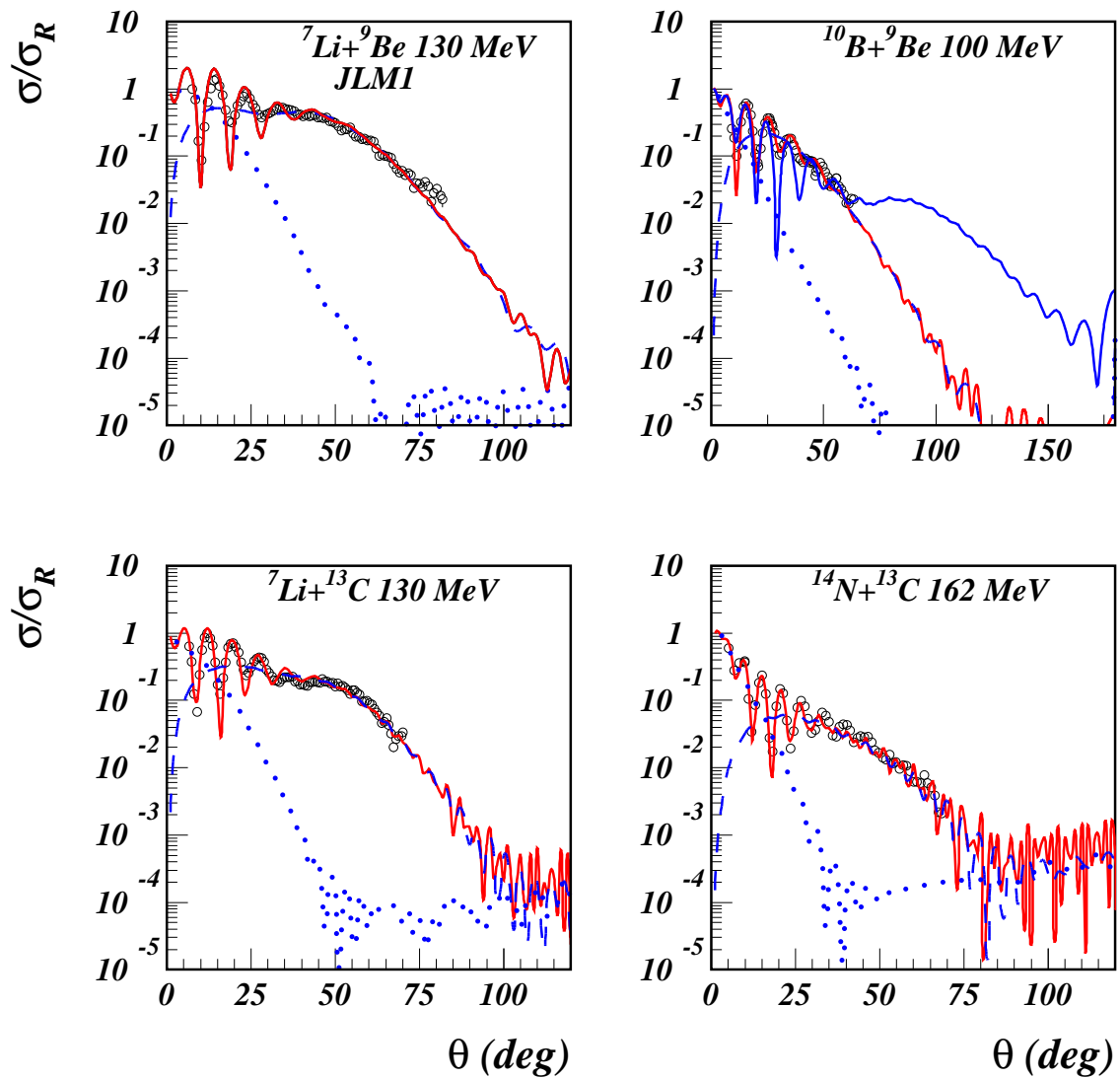,width=15cm}}
\end{center}
\caption{(Color online) Comparison of JLM1 folding model calculations with $%
^{7}$Li scattering data at 19 MeV/nucleon. The data are taken from 
\protect\cite{trache00}. Two JLM1 solutions are indicated for $^{10}$B+$^{9}$%
Be reaction. The parameters are given in Table \protect\ref{tab1:res_JLM1}.
Far-side (dashed) and near-side (dotted) cross sections are indicated in
ratio to Rutherford cross sections.}
\label{fig:jlm1_2}
\end{figure*}
As these give very similar results only JLM1 parameters are listed in Table %
\ref{tab1:res_JLM1} and the results of the calculations are shown in Figs. %
\ref{fig:jlm1_1} to \ref{fig:jlm1_3}. At 9 MeV/nucleon (Fig. \ref{fig:jlm1_1}%
) the same pattern emerges as with Woods-Saxon form factors. The pre-rainbow
oscillation is carried entirely by the dominant far-side component. Some
other high order structures appear at angles near 180$^{\circ }$ as the
result of near/far amplitude interference. At most forward angles this
interference produces an inner Fraunhofer crossing which give rise to a deep
minimum in the cross section.

For $^{7}$Li+$^{9}$Be at 63 MeV, JLM1 calculation failed to describe the
oscillation near $\theta =80^{\circ }$ for the simple reason that data
required a $rms$ radius for the real potential $R_{V}=3.4$ fm, while the
bare JLM interaction predicts a minimal $R_{V}$=3.6 fm for $t_{R}\approx 0$.
This once again reflects the critical role played by the radial behavior of
DPP. This is also illustrated in the upper right quadrant of Fig. \ref%
{fig:jlm1_2} where two JLM solutions for the reaction $^{10}$B+$^{9}$Be at
10 MeV/nucleon are indicated (see also Table \ref{tab1:res_JLM1}). The
solution with smaller real volume integral which better fits the forward
angles predicts a smooth, exponentially decaying cross section beyond $%
\theta \approx 60^{\circ }$. %fig 6
\begin{figure*}[tbh]
\begin{center}
\mbox{\epsfig{file=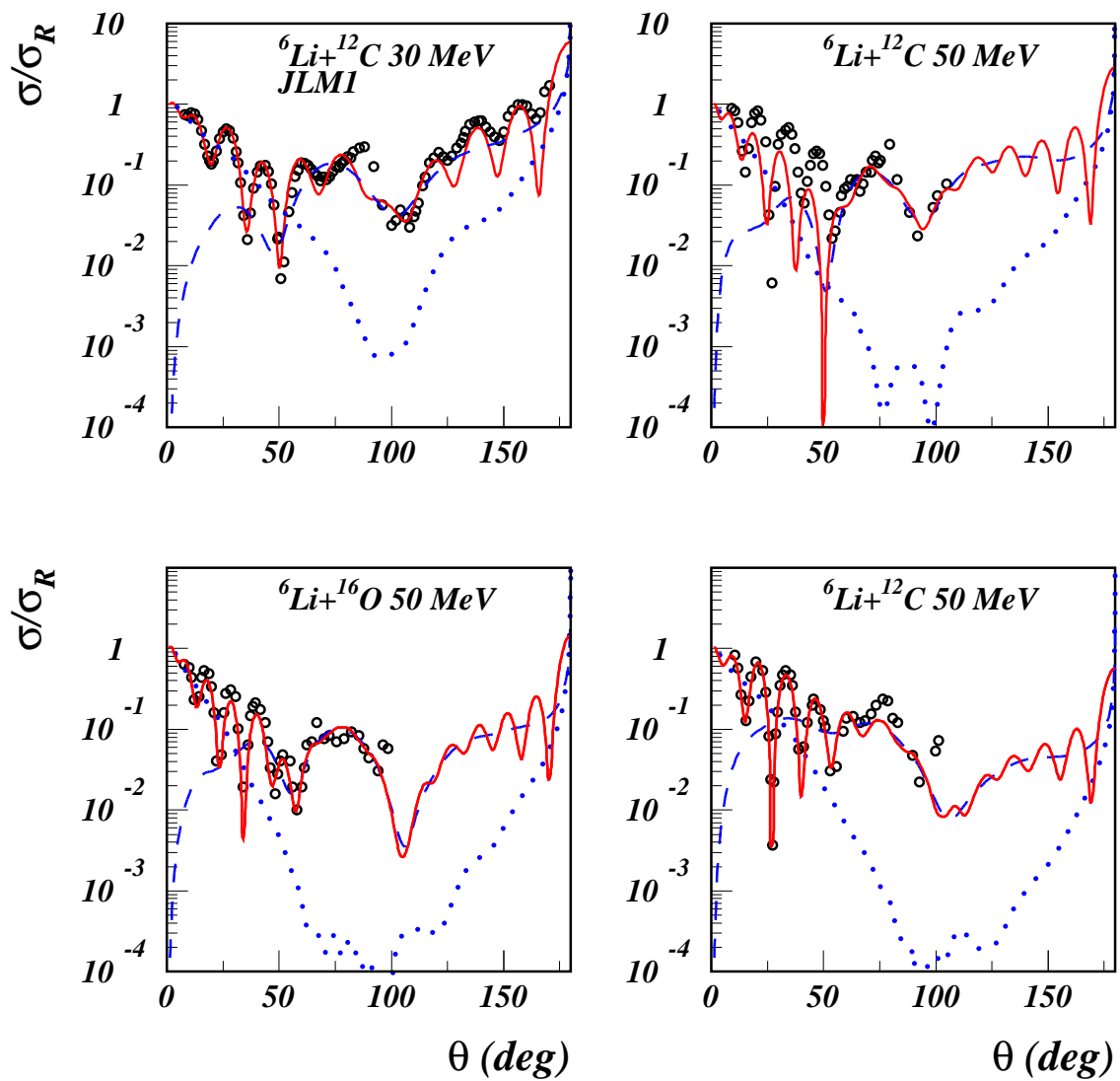,width=15cm}}
\end{center}
\caption{(Color online) Comparison of the JLM1 folding model calculations
with $^{6}$Li scattering data on light targets at 30 and 50 MeV laboratory
energy. For the $^{12}$C target at 50 MeV, only a solution with a real
volume integral exceeding the critical value $J_{V}$=300 MeV fm$^{3}$ (Table 
\protect\ref{tab1:res_JLM1}) is able to reproduce both forward and
intermediate angles (right bottom panel). Far-side (dashed lines) and
near-side (dotted) cross sections are indicated in ratio to Rutherford cross
sections. }
\label{fig:jlm1_4}
\end{figure*}
The second solution with a real volume integral close to the critical value $%
J_{Vcrit}\approx 300$ MeVfm$^{3}$ gives rise to a shallow pre-rainbow
oscillation at these angles (not covered by experiments). The high
selectivity of the pre-rainbow oscillations to the optical potentials is
also illustrated in Fig. \ref{fig:jlm1_4} where other $^{6}$Li scattering
data from literature, at somewhat lower energies, are explored. The $^{6}$Li+%
$^{12}$C data at 50 MeV \cite{trcka} could be described in the whole angular
range only with potentials exceeding the critical value of the real volume
integral found before. In Fig. \ref{fig:jlm1_3} we show $^{6}$Li+$^{12}$C
elastic scattering data at 7 energies between 15 and 50 MeV/nucleon. 
%implaying that
%this is perhaps a common feature of light ion interaction at 8-10 %%MeV/nucleon.
Now, even at high energy (Fig. \ref{fig:jlm1_3}) the JLM1 description of the
rainbow patterns is exemplary (to be compared with Figs. 6 a) and c) of Ref. 
\cite{trache00}). This suggests that the geometrical details of the optical
potential rather than the density dependence are essential for a correct
description of $^{6,7}$Li elastic scattering at low and intermediate
energies.

%fig 7 
\begin{figure}[tb]
\begin{center}
\mbox{\epsfig{file=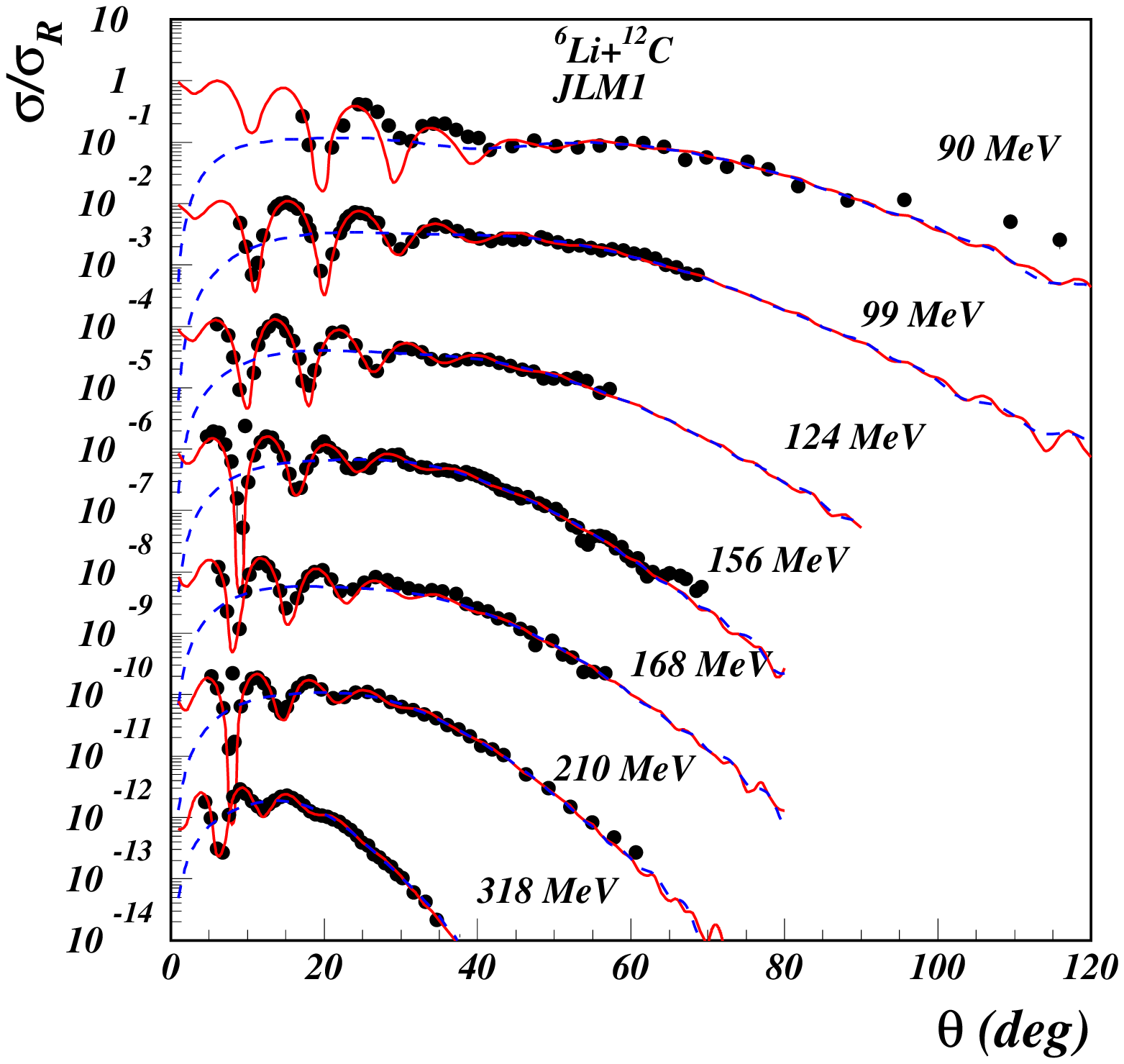,width=8.cm}}
\end{center}
\caption{(Color online) Comparison of JLM1 folding model calculations (full
lines) with high energy $^6$Li scattering data. The source of data and
calculation parameters are given in Table \protect\ref{tab1:res_JLM1}. The
far-side (dashed lines) cross sections are indicated in ratio to the
Rutherford cross sections (and from the top curve below, each case is
multiplied by an extra $10^{-2}$ factor). The near-side cross section (not
shown) is important only at most forward angles.}
\label{fig:jlm1_3}
\end{figure}

A close examination of the parameters in Table \ref{tab1:res_JLM1} reveals
an erratic variation of the range parameters $t_{V(W)}$ from one energy to
another and from system to system. As mentioned above, this largely reflects
the mass and energy dependence of DPP. 
% but can reflect also some inadequacy of our single
%particle densities.
The other parameters are more stable. 
%The strength  parameter $N_V$ increases slightly with energy and on average
% is somewhat larger that in our earlier analysis \cite{trache00}
%reflecting the need for stronger refractive effects,
%but again $N_W$ approach on average unity.
The strength parameter $N_{V}$ decreases slowly from 5 to 16.5 MeV/nucleon
and then increases again up to 53 MeV/nucleon, the highest energy at which
reliable data exist. This may suggest that DPP reaches its maximum amplitude
at energies around 16 MeV/nucleon. On average the $N_{V}$ values in Table %
\ref{tab1:res_JLM1} are somewhat larger than in our earlier analysis \cite%
{trache00} reflecting the need for stronger refractive effects, but again $%
N_{W}$ approaches unity, on average.

\section{Transfer reaction}

\label{sec4} As already mentioned, in one experimental run we have also
measured the neutron transfer reaction $^{13}$C($^{7}$Li,$^{8}$Li)$^{12}$C
at E($^{7}$Li)=63 MeV. The purpose of the study was to determine the ANC for
the ground state of $^{8}$Li, and then, using charge symmetry to relate it
with that in its mirror nucleus $^{8}$B. The ANC was then used to calculate
the astrophysical factor $S_{17}$ that gives the rate of the proton capture
reaction $^{7}$Be($p,\gamma $)$^{8}$B, of crucial importance for the solar
neutrino problem. The major advantage of the neutron transfer reaction over
its mirror proton transfer reaction is that it involves a stable beam, and,
therefore, a much more precise and detailed angular distribution could be
measured. That allowed the determination of the admixture of the minor
component $1p_{1/2}$ in the wave function of the ground state of $^{8}$Li
(and $^{8}$B, respectively), dominated by the $1p_{3/2}$ orbital. The
results of this experiment were reported in Ref. \cite{trache03}. In that
study we paid particular attention to the dependence of the results on the
optical model potentials used in the entrance and exit channels. % fig 9
\begin{figure}[tbh]
\begin{center}
\mbox{\epsfig{file=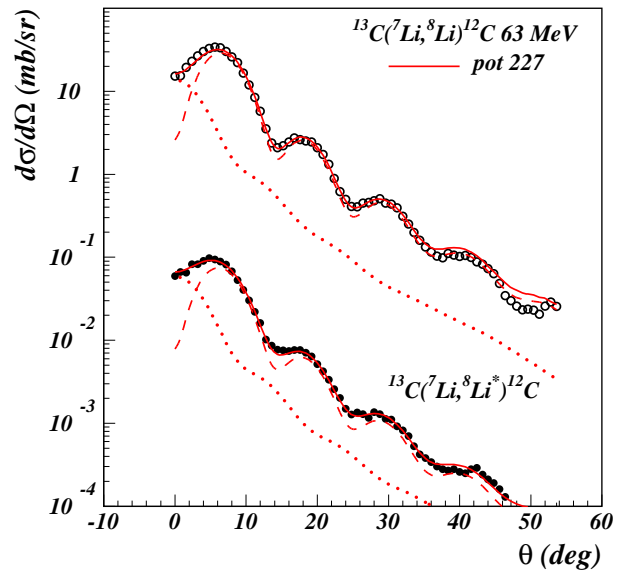,width=8.cm}}
\end{center}
\caption{(Color online) The angular distributions for the neutron transfer
reaction $^{13}$C($^{7}$Li,$^{8}$Li)$^{12}$C to the ground (top, open
points) and first excited state (bottom, full points) of $^{8}$Li. The
calculations shown (full line) are done using the potential "227" in Table 
\protect\ref{tab1:res_ws}. The data are shown as points and the separate
contributions of the $p_{1/2}\rightarrow p_{3/2}$(dashed line)and $%
p_{1/2}\rightarrow p_{3/2}$ (dotted line) components are shown in both cases.
}
\label{fig:transf}
\end{figure}

Eleven different combinations of entrance/exit potentials were used to show
that the resulting values for $C_{p_{3/2}}^{2}$ and $C_{p_{1/2}}^{2}$ are
very stable, when the potentials are reasonable. The potentials used were
either volume Woods-Saxon forms with the parameters from similar
projectile-target combinations at similar energies, or were obtained from
the double folding procedure with the renormalization coefficients from the
previous paper \cite{trache00}. Calculations done after the publication with
the new (deeper) potential "227" of Table \ref{tab1:res_JLM1} in both
entrance and exit channels lead to minor ($\sim $5\%) variations in the
results. The very good agreement between the experimental data and the DWBA
calculations and between the results of present and previous calculations
(Figure \ref{fig:transf}) shows that the region of the potential
contributing to transfer (the surface) is well and uniquely described. This
simultaneous description of elastic and transfer data is also an argument
for the complete determination of the optical potentials.

\section{Dispersion relation}

\label{sec5}The dispersion relation is a fundamental property of the optical
potential (see for example \cite{nagaraj}) and a selection between ambiguous
potentials can be performed by studying the dispersive properties of these
potentials, provided accurate analyses of experimental data are available
over a large energy range.

The threshold anomaly which manifests itself as a sharp increase of the real
optical potential for energies close to the Coulomb barrier, has been
explained by Nagarajan, Mahaux and Satchler \cite{nagaraj} as due to the
opening of reaction channels with increasing energy. An application of the
dispersion relation for elastic scattering of $^{16}$O on $^{208}$Pb at
energies around 80 MeV accounted well for this effect. Later it was
conjectured by Mahaux, Ngo and Satchler \cite{mahaux} that for loosely bound
nuclei, this anomaly may be absent. For these nuclei, the strong coupling
with breakup channels gives rise to a repulsive DPP which compensates the
strong attractive component. According to Sakuragi \cite{sakur} this effect
would explain the large renormalization needed by most of the effective
interactions used in the folding model for elastic scattering of $^{6,7}$Li.
The coupling with inelastic channels alone has been invoked by Gomez-Camacho 
\textit{et al.} \cite{gomez} to explain this reduction. The earlier analysis
of Kailas \cite{kailas} found strong dispersive effects for $^{6}$Li+$^{12}$%
C scattering. Recent studies by Tiede \textit{et al. }\cite{tiede} and by
Pakou \textit{et al.} \cite{pakou} of $^{6}$Li+$^{28}$Si at near barrier
energies found that the strength of the real part of the folding potential
using the M3Y interaction remains almost independent of energy, suggesting a
cancellation between the attractive (dispersive) component and the strong
repulsive dynamic polarization potential arising from the coupling to
continuum states. Another study of ~$^{6,7}$Li+$^{208}$Pb near the Coulomb
barrier \cite{keeley} found that at low energies the DPP is of opposite sign
for the two projectiles and there is a threshold anomaly for $^{7}$Li but
none for $^{6}$Li. No significant fusion hindrance caused by breakup effects
was found in the fusion reaction of $^{6,7}$Li on a $^{59}$Co target near
the Coulomb barrier \cite{beck}, thus leading the authors to conclude that
breakup suppression above the barrier appears to be a common feature of $%
^{6,7}$Li induced reactions.

\begin{figure}[tbh]
\begin{center}
\mbox{\epsfig{file=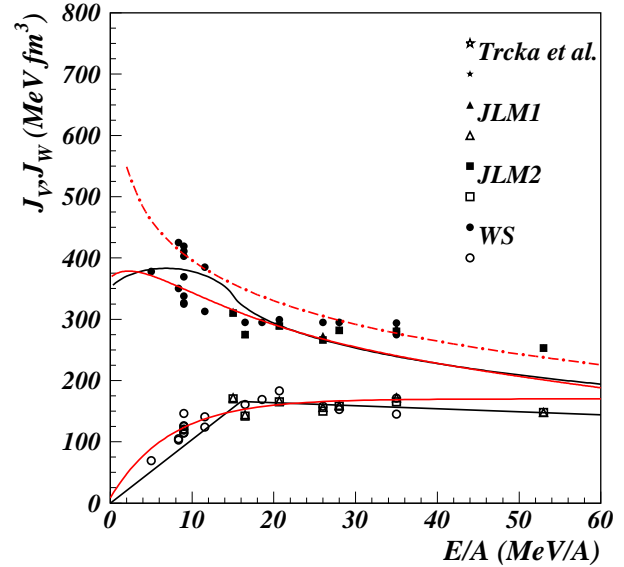,width=8.cm}}
\end{center}
\caption{(Color online) Energy dependence of the real (solid points) and
imaginary (open points) volume integrals obtained in the analyses with
Woods-Saxon and folding (JLM) optical potentials. The stars show the values
obtained from the OM1 optical potential of Ref. \protect\cite{trcka}. The
curves for $J_{V}$ are the result from the dispersion relation, normalized
to the empirical value at 26 AMeV, assuming the schematic models shown for $%
J_{W}$. The dash-dotted line gives the empirical energy dependence of the
real volume integral of Ref. \protect\cite{nadas3}.}
\label{fig:dispers}
\end{figure}

Therefore, the energy dependence of the $^{6,7}$Li optical potential is far
from clear and the competition between dispersive (attractive) and coupling
to continuum (repulsive) effects need to be studied more carefully. An
earlier study \cite{flomom} showed that the total reaction cross section for 
$^{6}$Li scattering saturates at energies around 20 MeV/nucleon and
therefore dispersive effects could be identified by accumulating good
optical potentials in this energy range. The real and imaginary volume
integrals for the optical potentials obtained in the previous sections are
plotted in Fig. \ref{fig:dispers}. Both Woods-Saxon and JLM folding results
have been included. These are supplemented with values derived from the
smooth OM1 potential of Trcka \textit{et al.} \cite{trcka}.

We assume that the local optical potential may be written as $V=V_{0}+\Delta
V(E)$ where $V_{0}$ is independent of energy and $\Delta V(E)$ is the energy
dependent DPP. We ignore the spurious energy dependence of $V_{0}$ arising
from non locality which is expected to be weak for heavy ions. We use the
dispersion relation connecting the imaginary and real volume integrals in
the subtracted form, 
\begin{equation}
J_{\Delta V,E_{s}}(E)=(E-E_{s})\frac{\mathcal{P}}{\pi }\int \frac{%
J_{W}(E^{\prime })}{(E^{\prime }-E_{s})(E^{\prime }-E)}dE^{\prime }
\label{eqdisp1}
\end{equation}%
where $E_{s}$ is a reference energy and $\mathcal{P}$ is the principal value
of the integral. In principle the evaluation of this equation requires the
knowledge of $J_{W}$ values at all energies. The above subtracted form takes
advantage of the fact that the energy dependence of $J_{W}$ far from
saturation energy is not very important and the unknown contributions are
absorbed by normalizing to the empirical value at a convenient reference
energy, 
\begin{equation}
J_{\Delta V,E_{s}}(E)=J_{\Delta V}(E)-J_{\Delta V}(E_{s})  \label{eqdisp2}
\end{equation}

Two schematic models have been employed here to estimate the energy
dependence of the imaginary volume integral. A first one approximates this
energy dependence by straight line segments \cite{mahaux}, which makes the
evaluation of Eq. \ref{eqdisp1} analytical. A more realistic energy
dependence is given by 
\begin{equation}
J_{W}(E)=J_{W}^{0}(1-\beta \exp (-\alpha E))  \label{eqdisp3}
\end{equation}%
where the parameters $J_{W}^{0}$=170 MeV fm$^{3}$, $\alpha $=0.023 MeV$^{-1}$
and $\beta $=0.95 describe better the energy dependence in the important
range 0-20 MeV/nucleon. In both calculations the reference energy was set to 
$E_{s}$=156 MeV, an energy where the JLM folding model gives precise values
for volume integrals. In general, the calculated dispersion contributions
get more repulsive as the energy increases, and the corresponding real
potentials get shallower, in qualitative consistency with phenomenology. An
empirical logarithmic dependence of the form $J_{V}=-785+95\ln (E)$ has been
found in Ref. \cite{nadas3} mostly based on unique OM potentials determined
from 35 and 53 MeV/nucleon $^{6}$Li scattering on light targets. This
matches perfectly the dependence obtained with the dispersion relation for E$%
>$10 MeV/nucleon, but disagrees at lower energies. In fact, this logarithmic
dependence is physically meaningful and can be understood on the basis of
the dispersion relation with a schematic (line segments) approach for the
imaginary volume integral. % fig 8

A relatively strong localized energy variation is predicted by the linear
model in the range 0-20 MeV/nucleon, while the exponential model predicts a
smooth dependence on the entire range of energies. This last calculation is
much closer to the data and seem to confirm $J_{V}$=320 MeV fm$^{3}$ as the
most realistic value at 9 MeV/nucleon, in surprising agreement with values
found for the more bound system $^{16}$O+$^{16}$O (see e.g. Fig 6 in ref. 
\cite{gonzales}). Most probably the phenomenological values found at 5
MeV/nucleon are due to the erratic variation in the WS parameters due to the
rapidly changing elastic scattering angular distributions \cite{vineyard}
near the resonance energy region around 20 MeV.

%%%%%%%%%%%%%%%%%%%%%%%%%%%%%

\section{Semiclassical barrier and internal barrier amplitudes}

\label{sec6} Once we have established the main features of the average OM
potential, we turn now to study the reaction mechanism in the elastic
scattering of $^{6,7}$Li on light targets at 9 MeV/nucleon using
semiclassical methods. The far-side dominance observed in the angular
distributions at 9 and 19 MeV/nucleon is not able to explain the differences
in the reaction mechanism at these energies. The reason is of course that
the far/near (F/N) decomposition method does not perform a dynamic
decomposition of the scattering function, but merely decomposes the
scattering amplitude into traveling waves. The intermediate angle
structures, such as those observed in our angular distributions, have been
repeatedly interpreted as arising from the interference of two ranges in
angular momenta $\ell _{<}$ and $\ell _{>}$ contributing to the same
negative deflection angle. However, the corresponding cross sections $\sigma
_{F<}$ and $\sigma _{F>}$ cannot be isolated because their dynamic content ($%
S$-matrix) is not accessible.

The semiclassical uniform approximation for the scattering amplitude of
Brink and Takigawa \cite{brink2} is well adapted to describe situations in
which the scattering is controlled by at most three active, isolated,
complex turning points. An approximate multireflection series expansion of
the scattering function can be obtained, the terms of which have the same
simple physical meaning as in the exact Debye expansion for the scattering
of light on a spherical well. The major interest in this theory comes from
the fact that it can give precious information on the response of a nuclear
system to the nuclear interior. Recent application \cite{anni} of this
technique helped to clarify the controversial problem of "Airy oscillation"
seen in low energy $^{16}$O+$^{12}$C scattering \cite{oglob}. % fig 10
\begin{figure}[tbh]
\begin{center}
\mbox{\epsfig{file=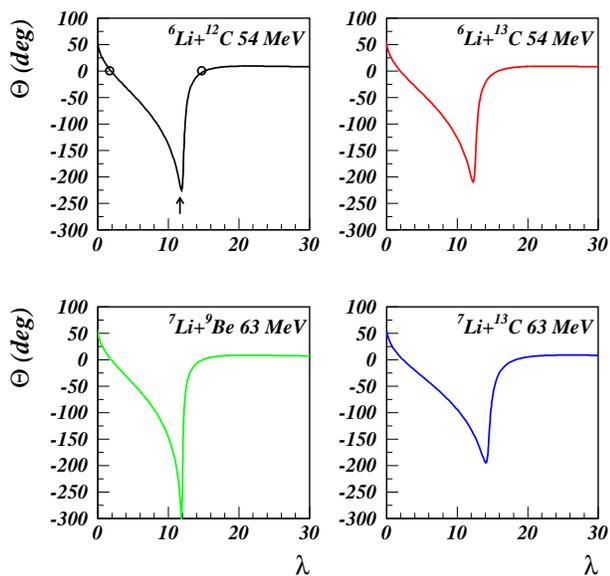,width=8.cm}}
\end{center}
\caption{(Color online) Semiclassical deflection functions for the real
potentials shown in Table \protect\ref{tab1:res_ws}. In the top left quad,
glories are indicated by open dots and the orbiting angular momentum by an
arrow.}
\label{fig:deflect}
\end{figure}

For the potentials in Table \ref{tab1:res_ws} (the first entry for each of
the four cases measured here) we discard the absorptive terms and define the
effective potential as, 
\begin{equation}
V_{eff}(r)=V(r)+\frac{\hbar ^{2}}{2\mu }\frac{\lambda ^{2}}{r^{2}},~~\lambda
=\ell +\frac{1}{2}  \label{eqsem1}
\end{equation}%
where the Langer prescription has been used for the centrifugal term. This
guarantees the correct behavior of the semiclassical wave function at the
origin \cite{frob}. Then we calculate the deflection function, 
\begin{equation}
\Theta (\lambda )=\pi -2\int_{r_{1}}^{\infty }\frac{\sqrt{\frac{\hbar ^{2}}{%
2\mu }}\lambda dr}{r^{2}\sqrt{E_{c.m.}-V_{eff}}}  \label{eqsem2}
\end{equation}%
where $r_{1}$ is the outer zero of the square root, i.e. the radius of
closest approach to the scatterer and $\mu $ is the reduced mass. Note that
with the replacement $\hbar \lambda =b\sqrt{2\mu E}$, Eq. \ref{eqsem2}
becomes identical with the classical deflection function $\Theta (b)$, where 
$b$ is the impact parameter. The results are shown in Fig. \ref{fig:deflect}%
. The behavior of $\Theta (\lambda )$ is the one expected for a strong
nuclear potential in a $\mathit{near~orbiting}$ kinematical situation in
which the c.m. energy approximately equals the top of the barrier for some
specific angular momentum. The deflection functions exhibit no genuine
minimum, but rather a pronounced cusp close to an orbiting logarithmic
singularity. Therefore any interpretation of structures in angular
distributions in terms of Airy oscillations can be discarded. Rather we need
an interpretation appropriate for orbiting, a well documented situation in
classical physics \cite{ford}. We identify the cusp angular momenta as
orbiting momenta ($\lambda _{o}$) since they are related with the
coalescence of two (barrier) turning points and the innermost turning point
given by the centrifugal barrier become classically accessible. There are
two branches that can be distinguished, an internal branch, for low active
momenta $\lambda <\lambda _{o}$ related to semiclassical trajectories which
penetrate into the nuclear pocket and a less developed external (barrier)
branch ($\lambda >\lambda _{o}$) related to trajectories deflected at the
diffuse edge of the potential. % fig 11
\begin{figure}[tbh]
\begin{center}
\mbox{\epsfig{file=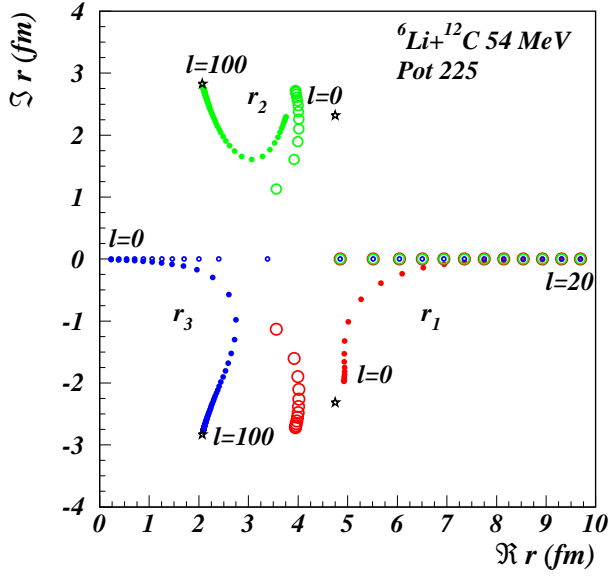,width=8.cm}}
\end{center}
\caption{(Color online) Complex turning points (full symbols) for the
potential "225" shown in Table \protect\ref{tab1:res_ws} at integer angular
momenta. Open symbols denote turning points for the real potential alone.
Stars indicate complex poles of the potential.}
\label{fig:turning}
\end{figure}

However this simple calculation cannot determine the relative importance of
these branches and provides no information about the interference effects of
the corresponding semiclassical trajectories. To clarify these points it is
best to go into the complex $r$-plane and look for complex turning points,
i.e. the complex roots of the quantity $E_{c.m.}-V_{eff}-iW$. This is an
intricate numerical problem, because, for a WS optical potential, the
turning points are located near the potential singularities and there are an
infinite number of such poles. The situation for integer angular momenta is
depicted in Fig. \ref{fig:turning} for the reaction $^{6}$Li+$^{12}$C at 54
MeV using the potential "225" in Table \ref{tab1:res_ws}. Only turning
points nearest the real axis are retained and we observe an ideal situation
with three, well isolated turning points for each partial wave. Even small
absorption plays an essential role in the motion of turning points. Removing
the imaginary part W, the barrier turning points ($r_{1,2}$) become complex
conjugates while the internal turning point is purely real (open symbols in
Fig. \ref{fig:turning}).

The multireflection expansion of the scattering function in the
Brink-Takigawa approach reads, 
\begin{equation}
S_{WKB}(\ell )=\sum_{q=0}^{\infty }S_{q}(\ell )  \label{eqsem3}
\end{equation}%
where, 
\begin{equation}
S_{0}(\ell )=\frac{\exp (2i\delta _{1}^{\ell })}{N(S_{21}/\pi )}
\label{eqsem4}
\end{equation}%
and for $q\not=0$, 
\begin{equation}
S_{q}(\ell )=(-)^{q+1}\frac{\exp {[2i(qS_{32}+S_{21}+\delta _{1}^{\ell })]}}{%
N^{q+1}(S_{21}/\pi )}  \label{eqsem5}
\end{equation}%
In these equations $\delta _{1}^{\ell }$ is the WKB (complex) phase shift
corresponding to the turning point $r_{1}$, $N(z)$ is the barrier
penetrability factor, 
\begin{equation}
N(z)=\frac{\sqrt{2\pi }}{\Gamma (z+\frac{1}{2})}\exp {(z\ln z-z)}
\label{eqsem6}
\end{equation}%
and $S_{ij}$ is the action integral calculated between turning points $r_{i}$
and $r_{j}$, 
\begin{equation}
S_{ij}=\int_{r_{i}}^{r_{j}}dr\{\frac{2\mu }{\hbar ^{2}}[E_{c.m.}-V_{eff}-iW]%
\}^{1/2}  \label{eqsem7}
\end{equation}%
$S_{21}$ and $S_{32}$ are independent of the integration path provided they
lie on the first Riemann sheet and collision with potential poles is
avoided. Each term in Eq. \ref{eqsem3} has a simple physical interpretation.
The first term (the barrier term, denoted also $S_{B}$) retains
contributions from trajectories reflected at the barrier, not penetrating
the internal region. The $q$th term corresponds to trajectories refracted $q$
times in the nuclear interior with $q$-1 reflections at the barrier turning
point $r_{2}$. Summation of terms $q\geq 1$ can be recast into a single
term, 
\begin{equation}
S_{I}=\frac{exp{[2i(S_{32}+S_{21}+\delta _{1}^{\ell })]}}{N(S_{21}/\pi )^{2}}%
\frac{1}{1+\exp {[2iS_{32}]/N(S_{21}/\pi )}}  \label{eqsem8}
\end{equation}%
and is known as the internal barrier scattering function. When the
absorption in the nuclear interior is large, the second factor in the above
equation reduces to one and we are left with the expression used in \cite%
{miche}. Since the semiclassical scattering function is decomposed
additively, $S_{WKB}=S_{B}+S_{I}$, the corresponding total scattering
amplitude is decomposed likewise as $f_{WKB}=f_{B}+f_{I}$ and conveniently
the corresponding barrier and internal barrier angular distributions are
calculated as $\sigma _{B,I}=|f_{B,I}|^{2}$, using the usual angular
momentum expansion of the amplitudes. % fig 12
\begin{figure}[tbh]
\begin{center}
\mbox{\epsfig{file=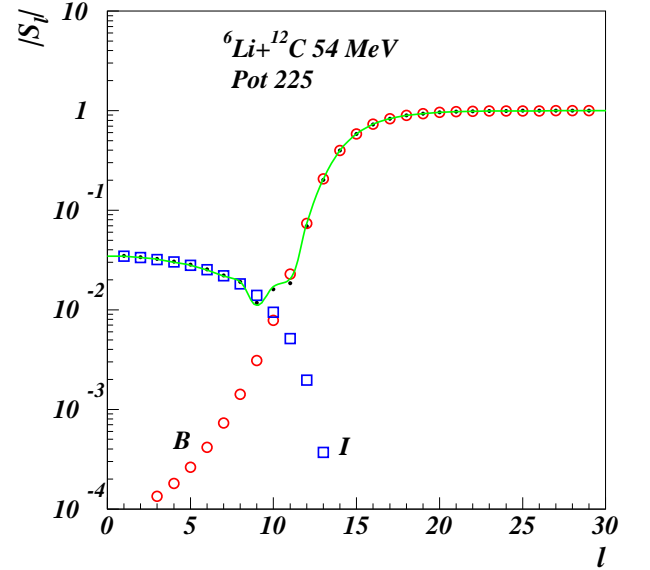,width=8.cm}}
\end{center}
\caption{ (Color online) Semiclassical decomposition of scatering function
for the WS potential of Fig. \protect\ref{fig:turning}. Barrier (open
circles) and internal barrier components (squares) are indicated. The exact
total quantum $S$-matrix is indicated by small dots. The line is a cubic
spline interpolation of the total semiclassical scattering function for the
same potential.}
\label{fig:profile}
\end{figure}

The accuracy of the semiclassical calculation has been checked by comparing
the barrier and internal barrier absorption profiles with the exact
quantum-mechanical result in Fig. \ref{fig:profile}. First, one observe that
the semiclassical B/I expansion is an $\mathit{exact}$ decomposition of the
quantum result. They are virtually identical at the scale of the figure. The
internal component gets significant values up to the grazing angular
momentum ($\ell _{g}$=15) and is negligibly small beyond this value. The
barrier component resembles a strong absorption profile and this justifies
the interpretation that it corresponds to that part of the flux not
penetrating into the nuclear interior. For values near the orbiting angular
momentum ($\ell _{o}\approx $12), the two components interfere and a
downward spike appears in the total profile, in complete agreement with the
quantum result. Second, the B/I components are almost decoupled in the
angular momentum space and therefore they will contribute in different
angular ranges. % fig 13
\begin{figure*}[tbh]
\begin{center}
\mbox{\epsfig{file=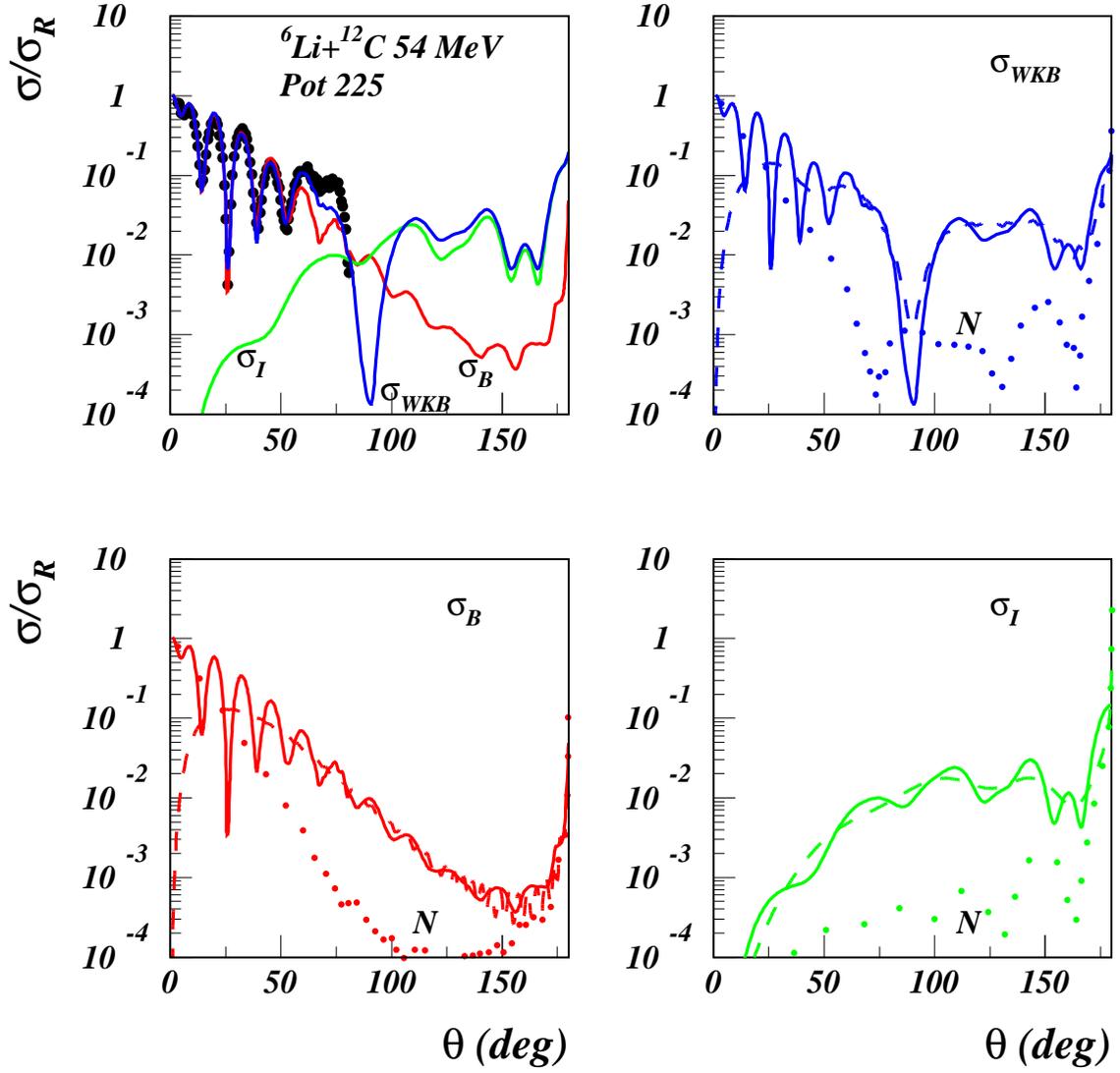,width=15cm}}
\end{center}
\caption{(Color online) Semiclassical barrier and internal barrier
decomposition of the cross section. The turning points and scattering
function are those from Figs. \protect\ref{fig:turning} and \protect\ref%
{fig:profile}, respectively. Each component is further dcomposed into
far-side (dashed) and near-side (dotted) components.}
\label{fig:wkbcross}
\end{figure*}

\begin{figure}[h]
\begin{center}
\mbox{\epsfig{file=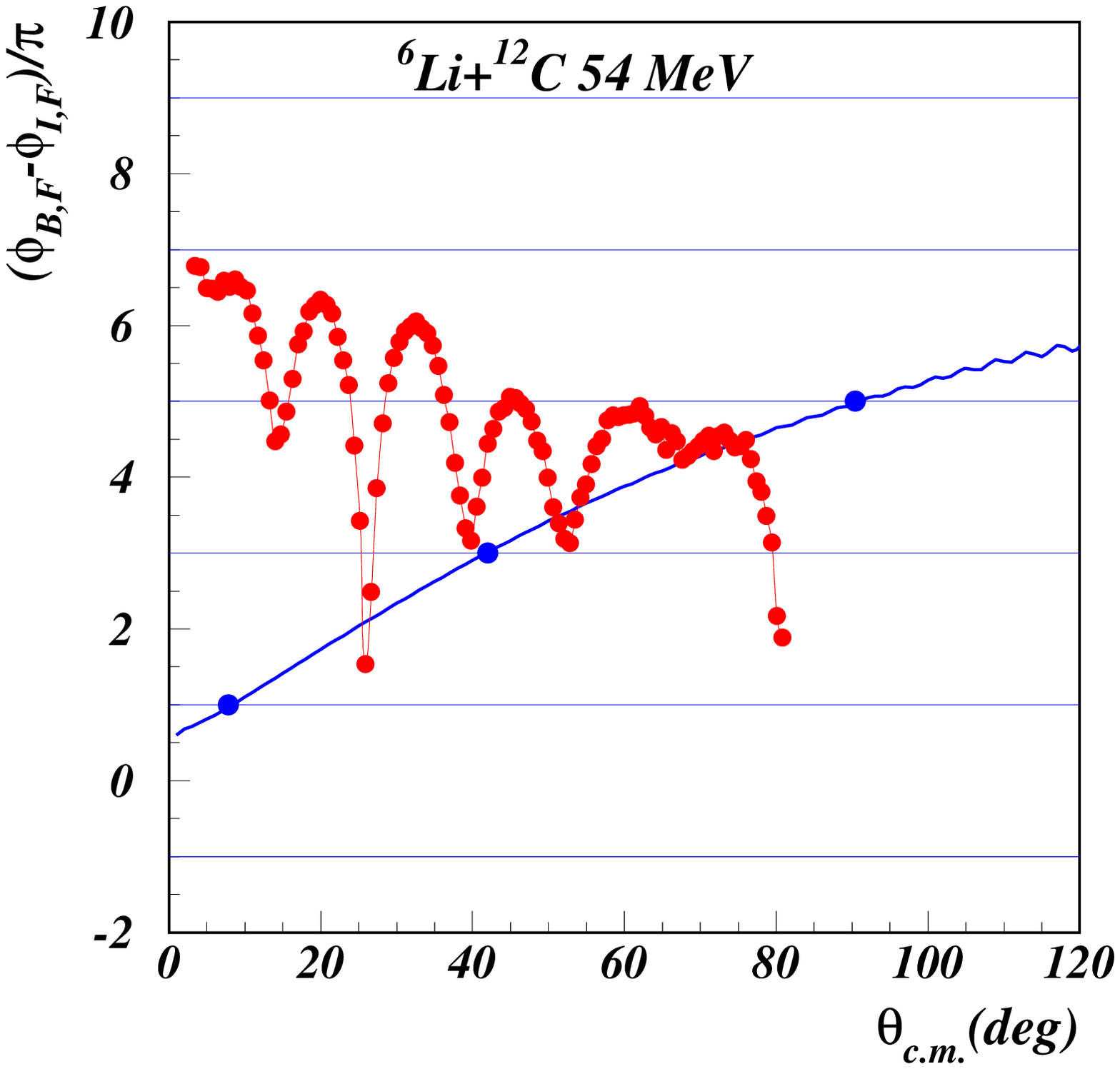,width=8.cm}}
\end{center}
\caption{ (Color online) The phase difference of the far-side
barrier/far-side internal barrier amplitudes as a function of scattering
angle. Large dots indicate the predicted interference minima. For easier
comparison, the experimental cross section is shown as $7+\log (\protect%
\sigma /\protect\sigma _{R})$ to match the scale of the figure. }
\label{fig:wkbmin}
\end{figure}

Semiclassical cross sections are compared with the data in Fig.\ref%
{fig:wkbcross} for the reaction $^{6}$Li+$^{12}$C at 54 MeV. Better insight
into this technique is obtained by further decomposing the B/I components
into far and near (BF/BN and IF/IN) subcomponents. Clearly, the barrier
component dominates the forward angle region. Fraunhofer diffractive
oscillations appear as the result of BF and BN interference. At large
angles, the internal contribution accounts for the full cross section. As
both B/I contributions are dominated by the far-side component (Fig. \ref%
{fig:wkbcross} bottom panels), we show in Fig. \ref{fig:wkbmin} the angles
at which the phase difference of the BF and IF amplitudes passes through an
odd multiple of $\pi $, i.e. where minima should be expected. Since the
crossing angle (where $\sigma _{B}\approx \sigma _{I}$) is about $\theta
\approx $ 75$^{\circ }$ and lies just in between predicted minima, the
coherent interference around this angle gives rise to the "plateau"
(constructive) and the deep minimum (destructive) at $\theta \approx $ 80$%
^{\circ }$. Similar consideration apply to the other three reactions. 
% fig 14

Thus, the intermediate angle exotic structure in angular distributions for
the elastic scattering of $^{6,7}$Li on light targets can be understood as a
result of coherent interference of two far-side subamplitudes generated by
different terms in the uniform multireflection expansion of the scattering
function (terms $q$=0 and $q$=1 in Eq. \ref{eqsem3}), corresponding to the
scattering at the barrier and the internal barrier. This interference effect
appears as a signature of a surprisingly transparent interaction potential
for loosely bound nuclei $^{6,7}$Li which allows part of the incident flux
to penetrate the nuclear interior and reemerge with significant probability.

\section{Conclusions}

\label{concl} We have performed precise measurements on extended angular
ranges of the elastic scattering of loosely bound nuclei $^{6,7}$Li on $%
^{12,13}$C and $^{9}$Be in four projectile-target combinations at 9
MeV/nucleon and reanalyzed previous data for the scattering of $^{7}$Li at
19 MeV/nucleon in an effort to obtain systematic information on the
interaction of $p$-shell nuclei with light targets. Optical potentials for
these nuclei are needed for studies in which highly peripheral transfer
reactions involving radioactive nuclei are used as indirect methods for
nuclear astrophysics and are an important factor in the accuracy and
reliability of these methods. At the present status of the experimental
techniques, the best information on the optical potentials for radioactive
nuclei can be obtained only by extrapolation from adjacent less exotic
nuclei. Our intention is to narrow the ambiguities in the optical model
potentials by systematic studies of the scattering of loosely bound
projectiles over a large range of angles and energies, and extract
information that can be used for systems involving radioactive projectiles,
for which elastic scattering data of very good quality are not easily
available. We demonstrate this procedure by reanalyzing the one neutron
transfer reaction $^{13}$C($^{7}$Li,$^{8}$Li)$^{12}$C using optical
potentials obtained in the present study.

The present data, which extend over a much larger angular range than
previously measured, confirm the existence of an exotic intermediate angle
structure, observed previously by Trcka \textit{et al.} It was interpreted
in Ref. \cite{trcka} as a diffractive effect arising from an angular
momentum dependent absorption. We adopt an opposite point of view and
interpret these structures as refractive effects arising from a fine balance
between the real and imaginary components of the optical potential. We have
performed a traditional analysis of our data in terms of Woods-Saxon and
microscopic JLM folded potentials. Both approaches lead to the conclusion
that the optical potential is deep and surprisingly transparent, in line
with findings for more bound systems. Folding model form factors have been
renormalized in the usual way in order to account for the energy and radial
dependence of the dynamic polarization potential. It is suggested that DPP
attains its maximum amplitude at approximately 16 MeV/nucleon for these
systems. The intermediate angle structures could be reproduced only with
potentials exceeding a critical volume integral of about 300 MeV fm$^{3}$
and, consequently, are severely selective, limiting the ambiguities in the
determination of the OMP. The remaining discrete ambiguities could be
removed by a dispersion relation analysis. Based on a good estimation of the
absorption at low energy (5-20 MeV/nucleon), this analysis allowed us to
extract a smooth energy dependence of the optical potential. Our analysis
did not find any spectacular anomaly near the Coulomb barrier and seems to
confirm, to some extent, the conjecture of a canceling effect between the
repulsive dynamic polarization potential due to the coupling with breakup
channels and the attractive, dispersive component of the optical potential.

In our previous study \cite{trache00} we found a simple recipe to obtain OMP
for loosely bound $p$-shell nuclei from a double folding procedure using the
JLM effective NN interaction. The already independent real and imaginary
parts were smeared with constant, but different ranges $t_{V}=1.2$ fm and $%
t_{W}=1.75$ fm, which accounted for the well known need for a wider
imaginary potential to describe the experimental data. We found that a
considerable renormalization of the real part was needed $N_{V}=0.37\pm 0.02$
(leading to volume integrals $J_{V}\simeq 220$ MeV fm$^{3}$), but not for
the imaginary part $N_{W}=1.00\pm 0.09$. That recipe was already
successfully applied to predict the elastic scattering angular distributions
of RNBs on light targets in a number of cases at energies around 10
MeV/nucleon. The present analysis shows that in order to reproduce the
structures observed at intermediate angles in the same cases measured, one
needs to\ allow for a more complicated radial dependence of the dynamic
polarization potential, energy and target dependent, and require deep real
potentials with volume integrals larger than a critical value of $%
J_{Vcrit}\approx 300$ MeV fm$^{3}$. This is a conclusion of the
phenomenological analyses and is supported by the dispersion relation
analysis. However, the elastic scattering data in the angular range of the
Fraunhofer oscillations and the transfer reactions can be equally well
described by the previous potentials produced by the folding procedure with
fixed smearing ranges for the effective NN interaction and the simple
renormalization of Ref. \cite{trache00}, showing that the potentials are
well described in the surface region.

In an effort to clarify the reaction mechanism responsible for the
intermediate angle structures found at 9 MeV/nucleon, we performed extensive
semiclassical calculations within the uniform multireflection expansion of
the scattering function of Brink and Takigawa. It has been shown that using
complex trajectories, the (external) barrier/internal barrier expansion is
an exact realization of the dynamic decomposition of the quantum result into
components responsible for that part of the incident flux reflected at the
barrier and the part of the flux which penetrates into the nuclear interior
and reemerges with significant probability. By combining the B/I
decomposition with the usual far-side/near-side expansion, we explain the
intermediate angle structure as a coherent interference effect of two
subamplitudes (BF and IF). Thus, this refractive effect appears as a
signature of a highly transparent interaction potential.

\begin{acknowledgements} 
This work was supported in part by the U. S. Department of Energy under 
Grant No. DE-FG03-93ER40773 and by the Robert A. Welch Foundation. One of 
the authors (F.C.) acknowledges the support of the Cyclotron 
Institute, Texas A\& M University and of the IN2P3 including that 
provided within the framework of the NIPNE-HH$-$IN2P3 convention.
\end{acknowledgements}

%%%%%%%%%%%%%%%%%%%%%%% biblio %%%%%%%%%%%%%%%%%%%%%%%%%%%%%%%%%%%%%%

%%%%%%%%%%%%%%%%%%%%%%%%%%%%%%%%%%%%%%%%%%%%%%%%%%%%%%%%%%%%%%%%%%%%
%
%\section{TABLES}
%
%%%%%%%%%%%%%%%%%%%%%%%%%%%%%%%%%%%%%%%%%%%%%%%%%%%%%%%%%%%%%%%%%%%%
%%
\begin{table}[tbp]
\caption{ List of the elastic scattering experiments presented in this
paper. }
\label{tabexp}%[tbh]
\par
\begin{center}
\begin{tabular}{cccc}
\hline\hline
&  &  &  \\ 
No. & Reaction & E [MeV] & $\theta _{lab}$[deg.] \\ 
&  &  &  \\ \hline
1 & $^{6}$Li + $^{12}$C & 54 & 2 - 56 \\ 
2 & $^{6}$Li + $^{13}$C & 54 & 2 - 59 \\ 
3 & $^{7}$Li + $^{9}$Be & 63 & 4 - 52 \\ 
4 & $^{7}$Li + $^{13}$C & 63 & 4 - 56 \\ 
5 & $^{7}$Li + $^{9}$Be & 130 & 4 - 47 \\ 
6 & $^{7}$Li + $^{13}$C & 130 & 4 - 47 \\ 
&  &  &  \\ \hline\hline
\end{tabular}%
\end{center}
\end{table}

%%%%%%%%%%%%%%%%%%%%%%%%%%%%%%%%%%%%%%%%%%%%%%%%%%%%%%
%%%%%%%%%%%% Table 2 : WS analysis summary
%%%%%%%%%%%%%%%%%%%%%%%%%%%%%%%%%%%%%%%%%%%%%%%%%%%%%%

\begin{table*}[tbh]
\caption{Best fit Woods-Saxon parameters. Reduced radii are defined in the
heavy ion convention. All lengths are given in fm, depths and energies in
MeV, cross sections in mb and volume integrals in MeV~fm$^3$. Coulomb
reduced radius is fixed to $r_c$=1~fm. $R_V$ and $R_W$ are the $rms$ radii
of the real and imaginary potentials, respectively.}
\label{tab1:res_ws}
\begin{center}
\begin{tabular}{cccccccccccccc}
\hline\hline
Reaction & Energy & $V_0$ & $W_0$ & $r_V$ & $r_W$ & $a_V$ & $a_W$ & $\chi^2$
& $\sigma_R$ & $J_V$ & $R_V$ & $J_W$ & $R_W$ \\ 
& [MeV] & [MeV] & [MeV] & [fm] & [fm] & [fm] & [fm] &  & [mb] & [MeV fm$^3$]
& [fm] & [MeV fm$^3$] & [fm] \\ \hline
$^6$Li+$^{12}$C & 54 & 225.47 & 15.75 & 0.503 & 1.157 & 0.900 & 0.737 & 17.71
& 1309 & 338 & 3.70 & 121 & 4.59 \\ 
&  & 371.31 & 17.70 & 0.439 & 1.109 & 0.856 & 0.777 & 13.60 & 1322 & 419 & 
3.47 & 125 & 4.56 \\ 
$^6$Li+$^{13}$C & 54 & 225.28 & 14.75 & 0.502 & 1.181 & 0.916 & 0.707 & 14.62
& 1327 & 327 & 3.76 & 114 & 4.63 \\ 
&  & 364.46 & 16.95 & 0.443 & 1.133 & 0.871 & 0.744 & 14.24 & 1338 & 403 & 
3.53 & 119 & 4.58 \\ 
$^7$Li+$^{9}$Be & 63 & 225.85 & 24.74 & 0.536 & 0.941 & 0.828 & 0.980 & 10.14
& 1456 & 369 & 3.49 & 146 & 4.66 \\ 
&  & 368.34 & 29.38 & 0.478 & 0.882 & 0.790 & 1.004 & 11.85 & 1470 & 464 & 
3.28 & 153 & 4.62 \\ 
$^7$Li+$^{13}$C & 63 & 227.94 & 15.37 & 0.529 & 1.186 & 0.932 & 0.669 & 20.09
& 1367 & 328 & 3.87 & 107 & 4.64 \\ 
&  & 278.86 & 24.19 & 0.594 & 1.050 & 0.789 & 0.721 & 20.04 & 1334 & 411 & 
3.53 & 126 & 4.38 \\ 
$^7$Li+$^{13}$C & 130 & 149.11 & 29.73 & 0.636 & 0.932 & 0.885 & 0.929 & 2.61%
$^a $ & 1403 & 282 & 3.90 & 132 & 4.62 \\ 
$^7$Li+$^{9}$Be & 130 & 143.41 & 33.64 & 0.581 & 0.829 & 0.892 & 1.094 & 3.03%
$^a$ & 1446 & 295 & 3.76 & 169 & 4.80 \\ 
$^{13}$C+$^{9}$Be & 130 & 159.85 & 24.43 & 0.674 & 0.983 & 0.868 & 0.914 & 
13.69 & 1552 & 280 & 3.96 & 104 & 4.79 \\ 
&  &  &  &  &  &  &  &  &  &  &  &  &  \\ \hline\hline
\end{tabular}%
\end{center}
\par
{\footnotesize $^a$ uniform 10\% errors. }
\end{table*}
%\newpage %%%%%%%%%%%%%%%%%%%%%%%%%%%%%%%%%%%%%%%%%%%%%%%%%%%%%%
%%%%%%%%%%%% Table 3 : JLM1 analysis summary
%%%%%%%%%%%%%%%%%%%%%%%%%%%%%%%%%%%%%%%%%%%%%%%%%%%%%%

\begin{table*}[tbp]
\caption{Best fit JLM1 parameters. The notations are those from the text.
Lengths are given in fm, energies in MeV, cross sections in mb and volume
integrals in MeV~fm$^3$. }
\label{tab1:res_JLM1}
\begin{center}
\begin{tabular}{cccccccccccc}
\hline\hline
Reaction & Energy & $t_V$ & $t_W$ & $N_V$ & $N_W$ & $\chi^2$ & $\sigma_R$ & $%
J_V$ & $R_V$ & $J_W$ & $R_W$ \\ 
& [MeV] & [fm] & [fm] &  &  &  & [mb] & [MeV fm$^3$] & [fm] & [MeV fm$^3$] & 
[fm] \\ \hline
$^6$Li+$^{12}$C & 30$^a$ & 0.30 & 2.45 & 0.60 & 0.46 & 14.8 & 1371 & 396 & 
3.66 & 72 & 4.93 \\ 
& 50$^h$ & 0.08 & 2.78 & 0.56 & 0.78 & 12.0 & 1315 & 373 & 3.64 & 120 & 4.42
\\ 
& 54 & 0.08 & 2.76 & 0.54 & 0.77 & 21.4 & 1556 & 351 & 3.64 & 116 & 5.11 \\ 
& 90$^b$ & 0.70 & 2.70 & 0.52 & 1.24 & 18.4 & 1591 & 313 & 3.73 & 173 & 4.96
\\ 
& 99$^c$ & 0.60 & 1.75 & 0.47 & 1.01 & 4.21 & 1225 & 277 & 3.69 & 145 & 4.27
\\ 
& 124$^d$ & 0.60 & 1.75 & 0.51 & 1.09 & 3.96 & 1243 & 292 & 3.69 & 168 & 4.28
\\ 
& 156$^e$ & 0.50 & 1.50 & 0.50 & 0.94 & 7.98 & 1146 & 271 & 3.66 & 154 & 4.19
\\ 
& 168$^d$ & 0.60 & 1.75 & 0.58 & 1.11 & 5.87 & 1231 & 305 & 3.68 & 185 & 4.28
\\ 
& 210$^f$ & 0.20 & 1.35 & 0.56 & 0.93 & 23.5 & 1062 & 276 & 3.59 & 161 & 4.05
\\ 
& 318$^g$ & 0.80 & 1.95 & 0.60 & 0.85 & 9.00 & 1069 & 251 & 3.69 & 148 & 4.35
\\ 
$^6$Li+$^{13}$C & 54 & 0.08 & 2.76 & 0.54 & 0.77 & 21.4 & 1556 & 351 & 3.64
& 116 & 5.11 \\ 
$^6$Li+$^{16}$O & 50$^h$ & 0.50 & 2.81 & 0.55 & 0.60 & 13.4 & 1643 & 346 & 
3.64 & 91 & 5.24 \\ 
$^7$Li+$^{9}$Be & 63 & 0.09 & 1.20 & 0.46 & 0.98 & 19.5 & 1538 & 274 & 3.64
& 152 & 4.80 \\ 
$^7$Li+$^{13}$C & 63 & 0.12 & 2.59 & 0.52 & 0.78 & 19.0 & 1652 & 335 & 3.74
& 113 & 5.07 \\ 
$^7$Li+$^{13}$C & 130$^i$ & 0.13 & 1.97 & 0.48 & 1.02 & 4.58 & 1392 & 280 & 
3.73 & 146 & 4.50 \\ 
$^7$Li+$^{9}$Be & 130$^i$ & 0.12 & 2.34 & 0.50 & 1.23 & 7.98 & 1404 & 304 & 
3.62 & 183 & 4.65 \\ 
$^{14}$N+$^{13}$C & 162 & 1.44 & 1.82 & 0.39 & 0.73 & 33.1 & 1563 & 220 & 
4.29 & 89 & 4.66 \\ 
$^{10}$B+$^{9}$Be & 100 & 1.89 & 1.02 & 0.30 & 1.01 & 6.9 & 1266 & 185 & 4.33
& 146 & 4.08 \\ 
&  & 0.47 & 2.28 & 0.48 & 0.93 & 29.6 & 1558 & 298 & 3.75 & 133 & 4.79 \\ 
&  &  &  &  &  &  &  &  &  &  &  \\ \hline\hline
\end{tabular}%
\end{center}
\par
{\footnotesize data from $^a$ \cite{viney}, $^b$ \cite{gluch}, $^c$ \cite%
{stanley}, $^d$ \cite{katori}, $^e$ \cite{rebel},$^f$ \cite{nadas1}, $^g$ 
\cite{nadas2}, $^h$ \cite{trcka}, $^i$ uniform 10 $\%$ errors}.
\end{table*}

\end{document}